\documentclass[fleqn,usenatbib]{mnras}

\usepackage{newtxtext,newtxmath}

\usepackage[T1]{fontenc}

\DeclareRobustCommand{\VAN}[3]{#2}
\let\VANthebibliography\thebibliography
\def\thebibliography{\DeclareRobustCommand{\VAN}[3]{##3}\VANthebibliography}

\usepackage{graphicx}	
\usepackage{amsmath}	
\usepackage{ragged2e}

\usepackage[flushleft]{threeparttable}

\newcommand{\edited}[1]{{\color{black}#1}}

\title[Semi-Analytical Model for Radio Relics]{A Semi-Analytical Model for the Formation and Evolution of Radio Relics in Galaxy Clusters}

\author[Y. Zhou et al.]
{Yihao Zhou,$^{1,2}$\thanks{E-mail: yihaozhou20@gmail.com}
Haiguang Xu,$^{1,3}$\thanks{E-mail: hgxu@sjtu.edu.cn}
Zhenghao Zhu,$^{1,4}$
Yuanyuan Zhao,$^{1}$
Shida Fan,$^{1}$
\newauthor
Chenxi Shan,$^{1}$
Yongkai Zhu,$^{1}$
Lei Hao,$^{4,5}$
Li Ji,$^{6}$
Zhongli Zhang\,$^{4,7}$
and Xianzhong Zheng\,$^{6,8}$
\\
$^{1}$School of Physics and Astronomy, Shanghai Jiao Tong University, 800 Dongchuan Road, Shanghai 200240, China\\
$^{2}$School of Physics, Xi'an Jiaotong University, No. 28 West Xianning Road, Xi'an 710049, China\\
$^{3}$Shanghai Frontiers Science Center for Gravitational Wave Detection, 800 Dongchuan Road, Shanghai 200240, China\\
$^{4}$Shanghai Astronomical Observatory, Chinese Academy of Sciences, 80 Nandan Road, Shanghai 200030, China\\
$^{5}$Key Laboratory for Research in Galaxies and Cosmology, Shanghai Astronomical Observatory, Chinese Academy of Sciences, 80 Nandan Road, Shanghai 200030, China\\
$^{6}$Purple Mountain Observatory, Chinese Academy of Sciences, 10 Yuanhua Road, Nanjing 210023, China
\\
$^{7}$Key Laboratory of Radio Astronomy, Chinese Academy of Sciences, Nanjing 210008, China\\
$^{8}$School of Astronomy and Space Sciences, University of Science and Technology of China, Hefei 230026, China\\
}

\date{}

\pubyear{2021}

\begin{document}
\label{firstpage}
\pagerange{\pageref{firstpage}--\pageref{lastpage}}
\maketitle

\begin{abstract}
Radio relics are Mpc-sized synchrotron sources located in the peripheral regions of galaxy clusters. 
Models based on the diffuse shock acceleration (DSA) scenario have been widely accepted to explain the formation of radio relics. 
However, a critical challenge to these models is that most observed shocks seem too weak to generate detectable emission, unless fossil electrons, a population of mildly energetic electrons that have been accelerated previously, are included in the models. 
To address this issue, we present a new semi-analytical model to describe the formation and evolution of radio relics by incorporating fossil relativistic electrons into DSA theory, which is constrained by a sample of \edited{14} observed relics, and employ the Press–Schechter formalism to simulate the relics in a \edited{$20^{\circ} \times 20^{\circ}$} sky field at 50, 158, and 1400 MHz, respectively.
Results show that fossil electrons contribute significantly to the radio emission, which can generate radiation \edited{four} orders of magnitude brighter than that solely produced by thermal electrons at 158 MHz, and the power distribution of our simulated radio relic catalog can reconcile the observed $P_{1400}-M_{\mathrm{vir}}$ relation.
\edited{We predict that $7.1\%$ clusters with $M_{\mathrm{vir}} > 1.2\times 10^{14}\,\mathrm{M}_{\odot}$ would host relics at 158 MHz, which is consistent with the result of $10 \pm 6\%$ given by the LoTSS DR2.}
It is also found that radio relics are expected to cause severe foreground contamination in future EoR experiments, similar to that of radio halos. 
The possibility of AGN providing seed fossil relativistic electrons is evaluated by calculating the number of radio-loud AGNs that a shock is expected to encounter during its propagation.
\end{abstract}




\section{Introduction}
\label{section:introduction}
Radio relics appear as diffuse synchrotron radio sources in the peripheral regions of some merging or cool-core clusters, typically showing an elongated morphology with a linear extent and strong polarization up to the 30\% level in integrated linear measurement \citep{Feretti, Brunetti2014_CReReview, vanweeren_review}. 
It has been found that about 30\% of the observed relics appear in pairs in merging clusters, which are located on the opposite sides of the cluster core along the merger axis, such as the double relics in Abell 3667, Abell 3376, and ZWCl 2341.1+0000 \citep{Hindson2014_A3667,George2015_3367,Weeren2009_zwcl2341}.
The radio spectral index of a typical relic tends to stay constant along the length and steepen gradually along the width \citep{DiGennaro2018_Sausage,Ensslin1998_RR}, which is presumably the result of the decay of relativistic electrons behind the shock front. 
Since the first detection (the radio source 1253+275 in the Coma cluster; \citealt{Giovannini_1985_Coma, Giovannini_1991_Coma}), 50 radio relics have been detected in the range of 100 MHz to 1.5 GHz, and 25 of them have been confirmed spatially coincident with shocks caused by either a minor or a major merger \citep{Feretti}.
Different from radio halos, which are exclusively discovered in massive systems \citep{vanweeren_review}, radio relics are hosted by clusters with a wide range of mass.
The largest linear size (LLS) of a relic often exceeds 1 Mpc (e.g., A115, A1240, and CIZAJ2242.8+5301; \citealt{Botteon2016a_A115, Hoang2018,Storm2018}), and the LLS of the newly detected relic in CIG 0217+70 even reaches about 3.5 Mpc \citep{Hoang2021}, making it the most extended one ever found.

Among the models proposed to explain the origin of radio relics, the ones based on diffuse shock acceleration (DSA), which was introduced by \citet{Ensslin1998_RR} who assumed that electrons producing the synchrotron radio emission are accelerated by merger-induced shocks $in\ situ$, seems to be most persuasive because they can naturally explain the elongated morphology of the relics, the relic-merger connection and the co-existence of X-ray shocks and relics in, e.g., Abell 115 \citep{Botteon2016a_A115}, RX J0603.3+4212 \citep{Ogrean_2013_toothbrush} and 1E 0657-56 \citep{Shimwell2014_Bullet}). During a typical DSA process electrons can increase their energy via the first-order Fermi acceleration, i.e., electrons trapped in a shock zone gain significant energy through multiple reflections when they encounter moving magnetic inhomogeneities \citep{Fermi_1949} with a rate proportional to their energy, which yields a power-law and time-independent electron energy spectrum. Detailed descriptions of the DSA theory can be found in \citet{Blandford_1987_DSA_review,Drury_1983} and \citet{Malkov_2001}. Recently the term  ``standard DSA theory''  has been introduced by \citet{Botteon2020} to refer to the specific case in which only thermal electrons participate in the DSA process, which has been frequently applied in both observational and numerical studies. 

Although the scenario based on the DSA has been supported by increasing observational evidence, some problems remain unsolved. One of the most severe challenges is that within the frame of standard DSA theory, most observed shocks seem to be too weak to generate detectable radio emission. For example, in the study of a sample of 10 radio relics associated with X-ray shocks, \citet{Botteon2020} found that in order to reproduce the measured radio luminosities of the relics over 10\% of the shock energy is needed to be transferred to the electrons, if all the accelerated electrons are extracted from the thermal pool. Apparently this acceleration efficiency is unreasonably high, and can be mitigated only by introducing other mechanisms. 
\citet{Vazza2020} also used the standard DSA model to simulate radio relics at 150 MHz/1.4 GHz and found that the radio radiation are powerful enough to reconcile the observed $P_{\mathrm{1400}}-M_{\mathrm{vir}}$ relation, but the acceleration efficiency adopted in calculation ($\eta_{\mathrm{e}}=1\%$) is much higher than estimations from other works (e.g., \citealt{Hoeft2007_RRsimu, Kang2017}). 
Another long-standing challenge is how to explain the inconsistency between the X-ray Mach number $\mathcal{M}_{\mathrm{X}}$, which is obtained by measuring the differences of both X-ray surface brightness and X-ray gas temperature across the shock front, and the radio Mach number $\mathcal{M}_{\mathrm{radio}}$, which is estimated based on the relation between the shock strength and the radio spectral index assuming the standard DSA theory \citep{Brunetti2014_CReReview}. For most relics $\mathcal{M}_{\mathrm{X}} < \mathcal{M}_{\mathrm{radio}}$ is found \citep{vanweeren_review}.

Both challenges can be mitigated by adding the fossil relativistic electrons into the shock acceleration model, which is a population of preexisting mildly relativistic electrons that have been previously accelerated by, e.g., merger-induced shocks, active galactic nuclei (AGN), and/or turbulence and then experienced a period time of decay \citep{Kang2012b, Pinzke2013, Bonafede_2014, Shimwell2015, VanWeeren2017_AGN}. When the fossil relativistic electrons encounter a shock, they are advected into the shock from upstream. Thermal electrons, on the other hand, are injected isotropically from the environment \citep{Drury_1983}. 
Compared with thermal electrons, fossil relativistic electrons become sufficiently energetic after reacceleration and are capable of producing detectable emissions in a relatively broader band even in a relatively weak shock \citep{Kang_2002}. Furthermore, when fossil electrons are taken into account in the model, the calculated $\mathcal{M}_{\mathrm{radio}}$ tends to reflect more about the characteristics of fossil electrons than those of shocks (i.e., the dynamic properties of the intracluster medium (ICM); \citet{vanweeren_review}; see also Section~\ref{section:case_fossil}), which helps explain its deviation from $\mathcal{M}_{\mathrm{X}}$.
\edited{It is worth noting that besides the potential contribution of fossil electrons, the difference between $\mathcal{M}_{\mathrm{X}}$ and $\mathcal{M}_{\mathrm{radio}}$ might also come from the different parts of the underlying Mach number distribution they follow \citep{Hong2015}, or the fact that $\mathcal{M}_{X}$ is more sensitive to the relic's orientation than $\mathcal{M}_{\mathrm{radio}}$ \citep{Wittor2021_Mach_number}.
}

In this work we attempt to establish a new semi-analytical model to investigate the origin and evolution of radio relics by taking into account the contributions of fossil relativistic electrons in a shock propagation model. 
We also investigate the properties of the radio relics simulated in a \edited{$20^{\circ} \times 20^{\circ}$} sky patch and estimate their influence on the low frequency radio observations in the future. 
Our model is built upon the following three assumptions (a) only two clusters participate in each merger, (b) during each merger only one pair of merger shocks and up to one pair of radio relics are generated, and (c) fossil relativistic electrons are randomly distributed in the regions where the shocks are observed.

This paper is organized as follows. 
In Section~\ref{section:RR_ob_samples} we describe the sample of observed relics, which is built based on the data quoted from literature, used to constrain the semi-analytical model.
In Section~\ref{section:Method} we present the method to establish the semi-analytical model. 
In Section~\ref{section:results} we constrain the parameters in the model using the observed relic sample, simulate the radio relics population in the \edited{$20^{\circ} \times 20^{\circ}$} sky field, and study the properties of the simulated radio relics.
In Section~\ref{section:discussion} we discuss the possibility of AGNs acting as the source of seed relativistic electrons,
the influence of radio relics as a contaminating foreground component on the detection of the Epoch of Reionization (EoR) signals in future observation, as well as 
the contribution of Coulomb collision
as an energy loss mechanism for the relativistic electrons.
Finally, we summarize our results in Section~\ref{section:conclusion}. Throughout this work we adopt a flat $\Lambda$CDM cosmology with \edited{$H_{0}=100h\,\mathrm{km}\,\mathrm{s}^{-1}\, \mathrm{Mpc}^{-1}=71\,\mathrm{km}\,\mathrm{s}^{-1}\, \mathrm{Mpc}^{-1}$}, $\Omega_{\mathrm{m}} = 0.27$, $\Omega_{\Lambda}=0.73$, $\Omega_{b}=0.046$, $n_{s}=0.96$ and $\sigma_{8}=0.81$.

\section{Observation Sample of Radio Relics}
\label{section:RR_ob_samples}
In order to constrain the properties of fossil electrons in the model to be outlined in Section~\ref{section:Method}, we select a sample of well-studied radio relics, which satisfy the following criteria: (1) a shock is confirmed in the X-ray observations at the position of the relic, (2) the relic and the shock are tightly associated with each other, and (3) there are reliable X-ray and radio observation studies of the shock and the relics. 
\edited{
We list the properties of the selected radio relics along with those of their corresponding hosting clusters in Table~\ref{table:relics_samples}, and present the Mach numbers $\mathcal{M}_{\mathrm{X}}$ and $\mathcal{M}_{\mathrm{radio}}$ in Table~\ref{table:shock_samples}.
Within the samples we select, the well-known double radio relics in A3667 have integrated radio spectral indices $\lesssim 1$ \citep{Hindson2014_A3667}, which cannot be explained by DSA theory and makes it impossible to calculate the corresponding $\mathcal{M}_{\mathrm{radio}}$ in our model. This might come from the large observation error or the fact that turbulence in the post-shock region alleviates the radiation cooling \citep{Kang2017}. Therefore, we do not include these two radio relics in our following simulation.}

\begin{table*}

	\caption{Sample of the observed radio relics}
	\label{table:relics_samples}
	\begin{tabular}{cccccccccc} 
		\hline
		Name & Position  $^{\rm a}$ & Redshift & $M_{\mathrm{vir}}$ $^{\rm b}$ & $\log(P_{\nu})$ $^{\rm c}$  & $\nu$ $^{\rm d}$ & \edited{$r_{\mathrm{relic}}$} $^{\rm e}$  &  LLS$^{\rm f}$  & \edited{$\log(\mathcal{P}_{\mathrm{CRe}}/\mathcal{P}_{\mathrm{total}})$} $^{\rm g}$ & $\log(f_{\text{N,fo}}/f_{\text{N,th}})$ $^{\rm h}$\\
 --- & --- & --- & M$_{\odot}$ & W/Hz & MHz & Mpc & Mpc & --- & --- \\
		\hline
		A115       & N & 0.197  & $1.05\times 10^{15}$  & 25.81          & 1400 & 1.32 & 2.44    &\edited{$-4.02$}   & \edited{$+0.50$}  \\
A521       & SE & 0.253 & $1.71\times 10^{15}$   & 24.41          & 610 & 0.93 & 1.00   &\edited{$-6.51$}     & \edited{$-1.40$} \\
A1240     & N & 0.159  & \edited{${5.39\times 10^{14}}^{\dagger}$}    & 23.59          & 1400 & 0.70  & 0.65 & \edited{$-5.58$}  & \edited{$-0.57$}\\
A1240     & S & 0.159  & \edited{${5.39\times 10^{14}}^{\dagger}$}   & 23.81          & 1400 & 1.10  & 1.25   &\edited{$-4.63$}  & \edited{$+0.29$}\\
A2255      & NE & 0.081 & $7.90\times 10^{14}$  & 23.25          & 1400 & 0.90  & 0.70  &\edited{$-5.80$}  &\edited{$-0.78$}\\
A2744      & NE & 0.308 & $1.39\times 10^{15}$    & 23.42           & 1400 & 1.56  & 1.62  &  \edited{$-5.01$}     &\edited{$0.25$} \\
A3376     & E & 0.046  & $4.74\times 10^{14}$  & 23.79          & 1400 & 0.52 & 0.95 & \edited{$-5.88$}  & \edited{$-0.84$}\\
A3376     & W & 0.046  & $4.74\times 10^{14}$ & 23.88          & 1400 & 1.43 & 0.80  &\edited{$-4.40$}  &\edited{$0.00$}\\
A3667     & N & 0.056 & $8.00\times 10^{14}$  & 25.21          & 1400 & 2.05 & 1.86 &\edited{$--^{\rm i}$}  & \edited{$--^{\rm i}$}\\
A3667     & S & 0.056 & $8.00\times 10^{14}$ & 24.12          & 1400 & 1.36 & 1.30  & \edited{$--^{\rm i}$} & \edited{$--^{\rm i}$} \\
1E 0657-5655 & E & 0.296 & $1.81 \times 10^{15}$ & 22.20       & 2100  & 1.00 & 0.93  & \edited{$-7.39$} &\edited{$-2.07$} \\
ACT-CL J0102-4915   & NW    & 0.870   & $2.16\times 10^{15}$   & 24.56          & 610 & 1.10  & 0.56 &\edited{$-4.94$} & \edited{$0.00$}\\
CIZA J2242.8+5301 & N   & 0.192 & $2.50\times 10^{15}$ & 24.41          & 1400 & 1.57 & 1.70  &\edited{$-6.50$}  &\edited{$-1.98$}\\
CIZA J2242.8+5301 & S   & 0.192 & $2.50\times 10^{15}$  & 23.06         & 1400  & 1.06  & 1.45 & \edited{$-7.35$}  & \edited{$-1.88$} \\
RXCJ1314-2515  & W  & 0.247 & $9.70\times 10^{14}$  & 24.57          & 325 & 0.55 & 1.10  &\edited{$-6.89$} &\edited{$-2.03$}\\
RX J0603.3+4212 & N  & 0.225  & $1.00\times 10^{15}$    & 25.78          & 1400 & 1.00    & 1.87 &\edited{$-4.28$} &\edited{$+0.41$}\\ 
		\hline
	\end{tabular}

\begin{justify}
$^{\rm a}$ The position of the radio relics in the cluster.\\
$^{\rm b}$ 
\edited{The virial mass $M_{\mathrm{vir}}$ are quoted from \citet{PlanckCollaboration2014}, \citet{Botteon2020}, \citet{Botteon2020_A2255}, \citet{Jee_2015}, and \citet{Jee_2016}, with the exception of Abell 1240 marked with an $\dagger$, which is calculated based on the method presented in \citet{Zhu2016}.}\\
$^{\rm c,d,e,f}$ Radio powers $P_{\nu}$ and the corresponding frequencies $\nu$, distances of the relic to the cluster center $R_{\mathrm{relic}}$, and LLS \citep{Feretti,Botteon2020}.\\
$^{\rm g}$ Ratios of the pressure generated by fossil relativistic electrons $P_{\mathrm{CRe}}$ to the total pressure $P_{\mathrm{total}}$ (see Section \ref{section:case_fossil}).\\
$^{\rm h}$ Density ratios between fossil electrons $f_{\mathrm{N,fo}}$ and thermal electrons $f_{\mathrm{N,th}}$ at $p = p_{\mathrm{inj}}$ (see Section~\ref{section:case_fossil}).\\
\edited{$^{\rm i}$ The pressure and density ratio for the radio relics in Abell 3667 are not provided since their integrated radio spectral indices $\lesssim 1$ \citep{Hindson2014_A3667}, which cannot be explained by DSA theory and makes it impossible to calculate the corresponding $\mathcal{M}_{\mathrm{radio}}$ and do the following simulation based on our model.}
\end{justify}
\end{table*}

\begin{table*}
	\caption{Mach numbers of the observed radio relics}
	\label{table:shock_samples}
	\setlength{\tabcolsep}{13pt}
	\renewcommand{\arraystretch}{1.25}
	\begin{tabular}{cccccc} 
		\hline
		Name & Position  & $\mathcal{M}_{\mathrm{X}}$ &
$\mathcal{M}_{\mathrm{radio}}$ & Reference (X-ray) & Reference (radio) \\
		\hline
		A115       & N  & $1.87_{-0.4}^{+0.5}$    & \edited{${\sim 4.58}^{*}$}  & \citet{Botteon2016a} & \citet{Govoni2001} \\
A521       & SE  & $2.13_{-1.13}^{+1.13}$ & $2.33_{-0.04}^{+0.05}$  & \citet{Botteon2020}          &  \citet{Macario2013} \\
A1240     & N   & \edited{${\sim 2}^{*}$}    & $2.3_{-0.1}^{+0.1}$    & \citet{Hoang2018}          & \citet{Hoang2018}\\
A1240     & S   & \edited{${\sim 2}^{*}$}     & $2.4_{-0.1}^{+0.1}$    & \citet{Hoang2018}          & \citet{Hoang2018}\\
A2255      & NE  & $1.36_{-0.16}^{+0.16}$   & \edited{$2.77_{-0.35}^{+0.35}$}   & \citet{Akamatsu2017}          & \edited{\citet{Pizzo2009_A2255}} \\
A2744      & NE & $1.7_{-0.3}^{+0.5}$   & $2.05_{-0.19}^{+0.31}$      & \citet{Hattori2017}        &\citet{Pearce2017}   \\
A3376     & E   & $1.5_{-0.1}^{+0.1}$    & $2.53_{-0.23}^{+0.23}$ & \citet{Urdampilleta2018}           &\citet{George2015} \\
A3376     & W   & $2.94_{-0.6}^{+0.6}$   & $3.57_{-0.58}^{+0.58}$ &\citet{Akamatsu2012}          & \citet{George2015}   \\
A3667     & N  & $1.68_{-0.16}^{+0.16}$  &\edited{$--^{\rm \star}$} &\citet{Finoguenov2010}         &\citet{Hindson2014}\\
A3667     & S  & $1.75_{-0.13}^{+0.13}$ &\edited{$--^{\rm \star}$} & \citet{Akamatsu_and_Kawahara2013}         & \citet{Hindson2014} \\
1E 0657-5655 & E & $1.87_{-0.13}^{+0.16}$ & $2.01_{-0.14}^{+0.19}$ & \citet{Botteon2016a}        & \citet{Shimwell2015}  \\
ACT-CL J0102-4915    & NW    & $2.78_{-0.38}^{+0.63}$    & $2.53_{-0.41}^{+1.04}$   & \citet{Botteon2020}          & \citet{Botteon2020} \\
CIZA J2242.8+5301   & N  & $2.7_{-0.4}^{+0.7}$    & $4.58_{-0.19}^{+0.19}$ & \citet{Akamatsu2015}   & \citet{Storm2018}  \\
CIZA J2242.8+5301   & S  & $1.7_{-0.3}^{+0.4}$    & $1.9_{-0.08}^{+0.08}$  & \citet{Akamatsu2015}         & \citet{Hoang2017}  \\
RXCJ1314-2515   & W  & $1.7_{-0.28}^{+0.40}$   & $3.18_{-0.45}^{+0.87}$  & \citet{Botteon2020}         &\citet{George2017} \\
RX J0603.3+4212 & N  & $1.7_{-0.42}^{+0.41}$  & $3.78_{-0.2}^{+0.3}$   & \citet{Ogrean2013}         & \citet{Rajpurohit2017} \\ 
		\hline
	\end{tabular}%
\edited{
\begin{justify}
$^{\rm \ast}$ The uncertainties of $\mathcal{M}_{\mathrm{radio}}$ for A115 and $\mathcal{M}_{\mathrm{X}}$ for A1240 are not provided in the corresponding reference.\\
$^{\rm \star}$ The $\mathcal{M}_{\mathrm{radio}}$ are not given for the radio relics in Abell 3667 since their integrated radio spectral indices $\lesssim 1$ \citep{Hindson2014_A3667}, which cannot be explained by DSA theory and makes it impossible to calculate the corresponding $\mathcal{M}_{\mathrm{radio}}$.
\end{justify}}
\end{table*}

\section{Semi-Analytic Model}
\label{section:Method}

\subsection{Cluster Model}
\label{section:Cluster_model}

\subsubsection{Mass Function and Merger History}
\label{section:PS}
\edited{Closely following \citet{Li2019} and \citet{Vazza2020}}, we employ a standard method to simulate the mass function of galaxy clusters by adopting Press-Schechter formalism \citep{PStheory_1974} and the cold dark matter (CDM) models, according to which the number of clusters with virial mass between $M$ and  $M+dM$ per unit of comoving volume at redshift $z$ is 
\begin{equation}
\begin{gathered}
n(M, z) d M=\sqrt{\frac{2}{\pi}} \frac{\langle\rho\rangle}{M} \frac{\delta_{\text{c}}(z)}{\sigma^{2}(M)}\left|\frac{d \sigma(M)}{d M}\right|\\ \times \exp \left[-\frac{\delta_{\text{c}}^{2}(z)}{2 \sigma^{2}(M)}\right] d M,
\end{gathered}
\end{equation}
where $\left<\rho\right>$ is the average density of our universe at present, $\delta_{\mathrm{c}}(z)$ is the critical linear overdensity at redshift $z$, and $\sigma(M)$ is the current root-mean-square (rms) density fluctuation within a sphere enclosing a total mass of $M$. We choose to use the power-law expression of $\sigma(M)$ considering the mass range of typical clusters \citep{Sarazin2002,Randall_2002}
\begin{equation}
\sigma(M)=\sigma_{8}\left(\frac{M}{M_{8}}\right)^{-\alpha},
\end{equation}
where $\sigma_{8}$ is the rms density fluctuation on the scale of $8\,h^{-1}$Mpc, $M_{8} = \left(4\pi/3\right)\left(8\,h^{-1}\mathrm{Mpc}\right)^{3}\left<\rho\right>$ is the total mass enclosed in a sphere with a radius of $8\,h^{-1} \mathrm{Mpc}$, and $\alpha = (n+3)/6$ with $n=-7/5$ is related to the primordial power spectrum.
\edited{We use a maximum redshift cutoff $z_{\mathrm{max}} = 4$ with the interval $\Delta z = 0.01$. We assume this redshift range is large enough considering the furthest radio relic observed by now is at $z=0.870$ (ACT-CL J0102-4915; \citealt{Lindner_2014}) and our simulation results in Section~\ref{section:results} show all the detectable relics are located at $z<2$.}
We use the extended Press-Schechter theory outlined in \citet{ExtendedPS1993} to describe the merger history of clusters. Given its present mass and redshift for each cluster, Monte-Carlo simulation is run to trace the mass growth history by identifying merger events and generating a specific merger tree. We set the minimum mass change in a merger to be $\Delta M_{\mathrm{c}}=2 \times 10^{13}\,\text{M}_{\odot}$, thus a mass growth with $\Delta M \leq \Delta M_{\mathrm{c}}$ is treated as an accretion event rather than a merger. 

Now consider a cluster that has a mass of $M_{\mathrm{t2}}$ at time $t_{2}$. The probability that this cluster has a mass $M_{\mathrm{t1}}$ $(M_{\mathrm{t1}}<M_{\mathrm{t2}})$ at an earlier time $t_{1}$ $(t_{1} < t_{2})$ is 
\begin{equation}
\begin{gathered}
\operatorname{Pr}\left(M_{\text{t1}}, t_{1} \mid M_{\text{t2}}, t_{2}\right) d M_{\mathrm{t1}}=\frac{1}{\sqrt{2 \pi}} \frac{M_{\mathrm{t2}}}{M_{\mathrm{t1}}} \frac{\delta_{\text{c1}}-\delta_{\text{c2}}}{\left(\sigma_{1}^{2}-\sigma_{2}^{2}\right)^{3 / 2}}\\
\times\left|\frac{d \sigma_{1}^{2}}{d M_{\mathrm{t1}}}\right| \exp \left[-\frac{\left(\delta_{c 1}-\delta_{c 2}\right)^{2}}{2\left(\sigma_{1}^{2}-\sigma_{2}^{2}\right)}\right] d M_{t1},
\end{gathered}
\end{equation}
where $i=1,2$ is used to denote parameters at time $t_{1}$ and $t_{2}$, respectively. By introducing $\Delta S\equiv \sigma^{2}_{2}(M_{\mathrm{t2}}) - \sigma^{2}_{1}(M_{\mathrm{t1}})$ and $\Delta \omega \equiv \delta_{\mathrm{c2}}(t_2) - \delta_{\mathrm{c1}}(t_1)$, the equation can be simplified as 
\begin{equation}\label{equ:sim_extend_PS}
\operatorname{Pr}(\Delta S, \Delta \omega) d \Delta S=\frac{1}{\sqrt{2 \pi}} \frac{\Delta \omega}{(\Delta S)^{3 / 2}} \exp \left[-\frac{(\Delta \omega)^{2}}{2 \Delta S}\right] d \Delta S.
\end{equation}
\edited{Note that in equation~\ref{equ:sim_extend_PS}, $\Delta \omega$} should satisfy the condition
\begin{equation}
\Delta \omega \lesssim \Delta \omega_{\max }=\left[S\left|\frac{d \ln \sigma^{2}}{d \ln M}\right|\left(\frac{\Delta M_{c}}{M}\right)\right]^{1 / 2}
\end{equation}
in order to resolve the minimum mass change $\Delta M_{\mathrm{c}}$ \citep{ExtendedPS1993}; in this work we adopt $\Delta \omega=\Delta \omega_{\mathrm{max}}/2$ \citep{Randall_2002}. When $\Delta \omega$ is given, \edited{$\Delta S$} can be drawn randomly from a cumulative probability distribution of 
\begin{equation}
\begin{gathered}
\operatorname{Pr}(<\Delta S, \Delta \omega)=\int_{0}^{\Delta S} \operatorname{Pr}\left(\Delta S^{\prime}, \Delta \omega\right) d \Delta S^{\prime}\\
=1-\operatorname{erf}\left(\frac{\Delta \omega}{\sqrt{2 \Delta S}}\right),
\end{gathered}
\end{equation}
where $\operatorname{erf}(x)=(2 / \sqrt{\pi}) \int_{0}^{x} e^{-t^{2}} d t$ is the error function.

\begin{figure}
	\includegraphics[width=\columnwidth]{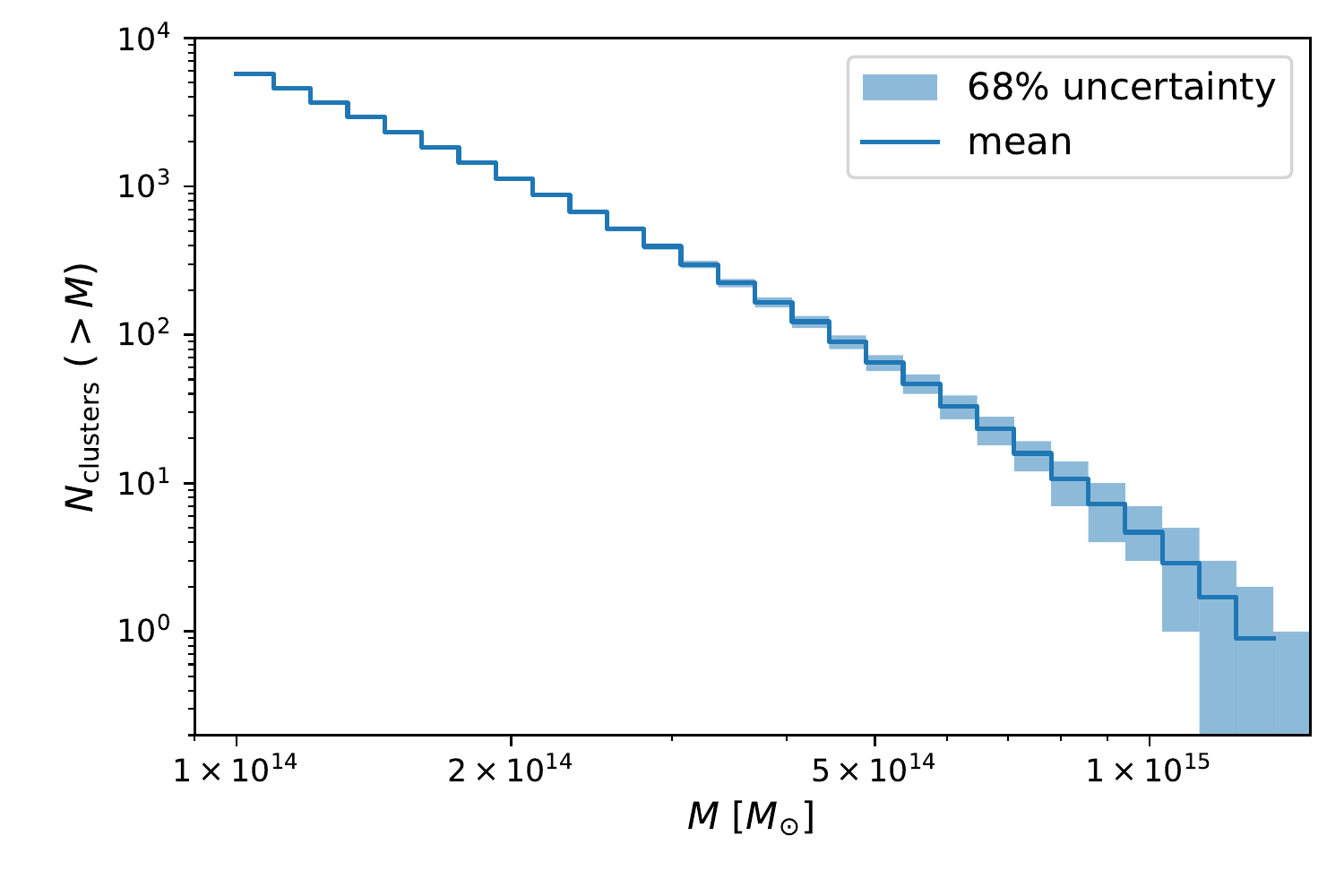}
    \caption{\edited{Cumulative number counts of clusters simulated in a $20^{\circ} \times 20^{\circ}$ sky field with the cut-off redshifts of $z=4$}. }
    \label{fig:massfunction}
\end{figure}

We limit our simulation to a \edited{$20^{\circ} \times 20^{\circ}$} sky patch, which is several times larger than the field of view \edited{(i.e., the amount of sky the telescope can image at once)} of the planned Square Kilometers Array (SKA) \citep{SKA_design}, in order to make sure that we have included sufficient number of clusters to carry out statistical analysis. 
We set the lower limit of cluster mass to be $M_{\mathrm{min}} = 1\times 10^{14}\, \mathrm{M}_{\odot}$ since less massive systems are found to be incapable of generating detectable signals (see Section~\ref{section:results}). 
The halo mass function is shown in Figure~\ref{fig:massfunction}, \edited{which is validated by the observed results given by \citet{Bohringer2017_massfunction}.}
Because radio relics emit through synchrotron radiation and their lifetimes ($\lesssim$ 0.1 Gyrs; \citealt{Brunetti2014_CReReview}) are short compared with the duration of typical merger events (about a few Gyr; e.g., \citealt{Hu_2021}), we trace the evolution of each cluster back 3 Gyr from the cluster's age at its redshift. 
Under these conditions we obtain \edited{8704} clusters in the simulation and \edited{5107} of them have experienced a merger.

\begin{figure}
	\includegraphics[width=\columnwidth]{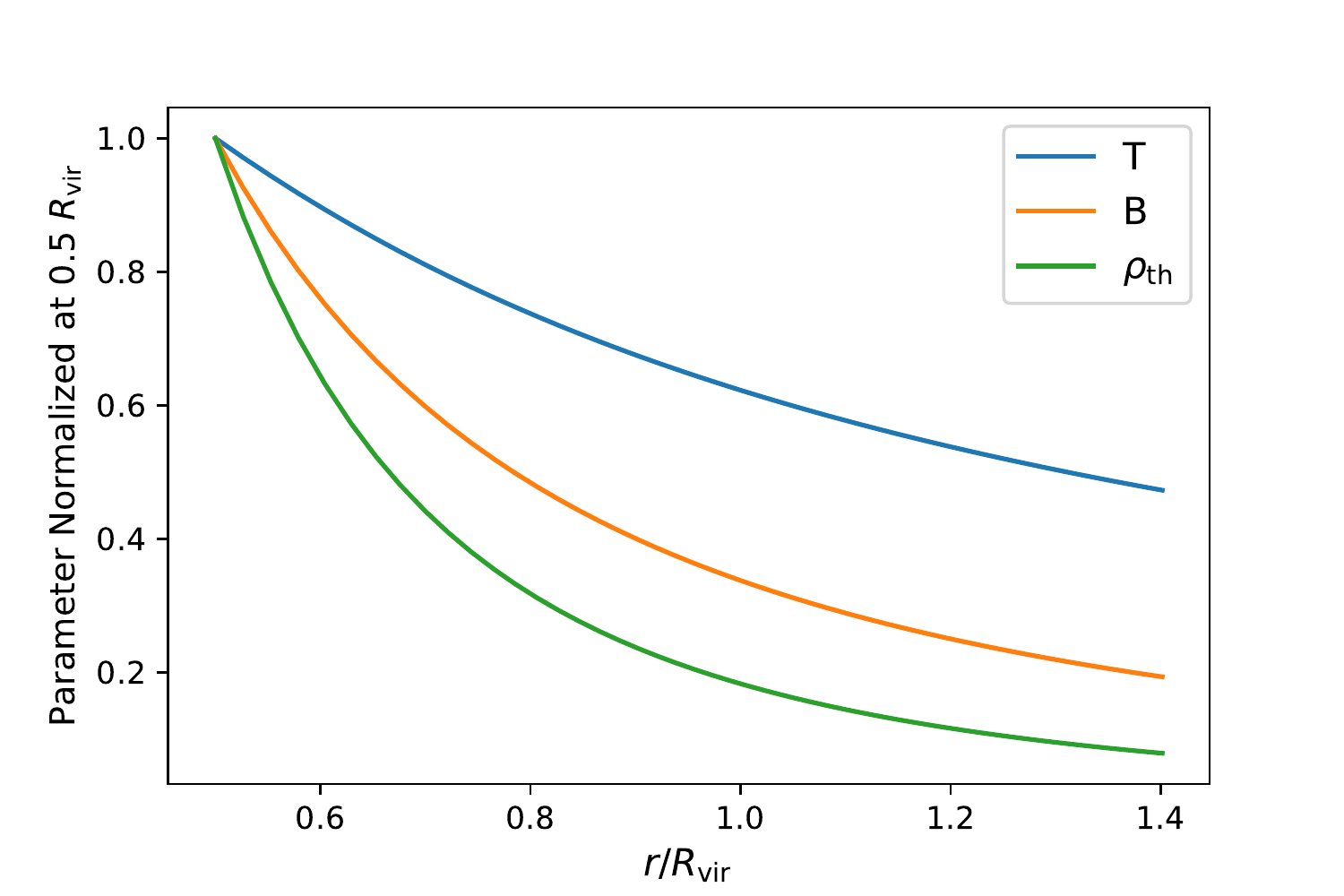}
    \caption{Normalized distribution of gas density $\rho_{th}$, gas temperature $T$ and magnetic field $B$. \label{fig:cluster_profile}}
\end{figure}

\subsubsection{Properties of ICM and Merger Shocks}
\label{section:cluster_para}
In our model the ICM mass density $\rho (r)$ is represented by a $\beta$ model \citep{Beta_model_Cava1976}
\begin{equation}\label{equ:beta_model}
\rho(r)=\rho(0)\left[1+\left(\frac{r}{r_{\mathrm{c}}}\right)^{2}\right]^{-3 \beta / 2},
\end{equation}
where $r_{\mathrm{c}}$ is the core radius and is fixed to $r_{\mathrm{c}} = 0.1R_{\mathrm{vir}}$ ($R_{\mathrm{vir}}$ is the virial radius of the cluster; \citealt{Sanderson2003}), and $\rho(0)$ is the central gas density.
\edited{The widely adopted value for the slope parameter $\beta$ is $2/3$ \citep{Jones1984, Li2019, Vazza2020}, which fits the gas density profile well at the central part of the clusters, while cannot give good description at $r>R_{\mathrm{500}}$ \footnote{$r_{500}$ is the characteristic radius of the cluster that the mean enclosed mass density is 500 times the critical density of the universe at the given redshift, and $M_{500}$ represents the total gravitating mass within $r_{\text{500}}$. Similarly, $M_{200}$ is the total gravitating mass within $r_{200}$.}, where the profile is steeper \citep{Vikhlinin_2006, Ghirardini_2019}. Since the radio relics are usually observed at cluster outskirts, the gas density profile in this region should be treated more carefully.
We set $\beta=5/6$ for $r\geq 0.5\,R_{\mathrm{200}}$, i.e., $\rho(r) \sim r^{-2/5}$ \citep{Zhang2019}, and $\beta=2/3$ for $r< 0.5\,R_{\mathrm{200}}$.
$\rho(0)$ can be constrained by the total ICM mass
$M_{\mathrm{gas}} = f_{\mathrm{gas}}M_{\mathrm{\mathrm{vir}}}$, where $f_{\mathrm{gas}}=\Omega_{b}/\Omega_{m}$.
Although this piecewise $\beta$ model is not differential at $0.5\,R_{\mathrm{200}}$, the shock begins to produce radio relics at $R_{\mathrm{sh,min}} = 0.5\,R_{\mathrm{200}}$ (see Section~\ref{section:shock_model}), i.e., we use $\beta=5/6$ for the whole process of shock evolution. We only require the gas density in the central part ($r < 0.5\,R_{\mathrm{200}}$) to calculate the $\rho(0)$.
}
The number density of the thermal electrons is then given by $n_{\mathrm{th,e}} = \rho \left(r\right)/\mu m_{\text{e}}$, where the mean molecular weight $\mu \sim 1.155$ \citep{Ettori2013_clustermassprofile} and $m_{\text{e}}$ is the electron mass. 

We assume that the gas temperature profiles in the clusters follow the form given in \citet{Vikhlinin_2006}, i.e.,  \begin{equation}\label{equ:T_profile}
\frac{T(r)}{T_{\mathrm{mg}}}=1.35 \frac{(x / 0.045)^{1.9}+0.45}{(x / 0.045)^{1.9}+1} \frac{1}{\left(1+(x / 0.6)^{2}\right)^{0.45}},
\end{equation}
where $x=r/r_{500}$, $T_{\mathrm{mg}}$ is the mass weighted average temperature for the whole cluster (except for the central region $r < 0.05\,R_{\mathrm{vir}}$), which scales with $M_{500}$
using the following relation \citep{Finoguenov2001_massTscale}
\begin{equation}\label{equ:M_T_scale}
M_{500}=\left(3.57_{-0.35}^{+0.41}\right) \times 10^{13} \times k T_{\mathrm{mg}}^{1.58_{-0.07}^{+0.06}}.
\end{equation}
Within the frame of the Navarro-Frenk-White (NFW) model, the total mass within the radius $r=s\,r_{\mathrm{vir}}$ is \citep{Lokas2001}
\begin{equation}
M\left(<s\,r_{\text {vir }}\right)=M_{\text {vir }} \frac{\ln (1+c s)-c s /(1+c s)}{\ln (1+c)-c /(1+c)},
\end{equation}
where $s$ can be seen as the radius in units of $r_{\mathrm{vir}}$.
The concentration parameter $c$ is related to the virial mass $M_{\mathrm{vir}}$, which can be approximated by $M_{200}$ \citep{Ettori2009}
\begin{equation}
c=A\left(\frac{M_{200}}{M_{\text {pivot }}}\right)^{B}(1+z)^{C},
\end{equation}
where $M_{\mathrm{pivot}}=2\times 10^{12}\ h^{-1}\text{M}_{\odot}$, $A=5.71$, $B=-0.084$ and $C=0.47$ \citep{Duffy2008}.

The magnetic field in the ICM is less constrained with present data or models. The magnetic field has been measured only for a few radio relics, and the results show that $B \sim \mu G$. In this work we adopt the approach of \citet{Beck2005_mag_equipartition} by assuming that the energy density of the magnetic field $\epsilon_{B}$ has reached equipartition with that of the cosmic rays (CR)
\begin{equation}\label{equ:B_profile}
\epsilon_{B} = \frac{B^{2}}{8\pi} = \chi_{\mathrm{cr}}\epsilon_{\mathrm{th}},
\end{equation}
where the coefficient $\chi_{\mathrm{cr}} = 0.015$ \citep{Li2019,Vazza2020} and $\epsilon_{\mathrm{th}}=3n_{\mathrm{th}}k_{B}T(r)/2$ is the thermal energy density of ICM. This yields a radial profile $B(r)$ that peaks at the cluster center and decays with radii, which is usually believed to be close to the lower limit of the true values \citep{Vazza2020}.
The normalized profiles of gas density, gas temperature, and magnetic field are shown in Figure~\ref{fig:cluster_profile}.

The jumps of the gas density and the temperature across the shock front are both described by the Rankine-Hugoniot relation
\begin{equation}\label{equ:RH}
\left\{\begin{array}{l}
\rho_{2}/\rho_{1}=4 \mathcal{M}_{\mathrm{X}}^{2}/(\mathcal{M}_{\mathrm{X}}^{2}+3),
\\
\\
T_{2}/T_{1}=(5 \mathcal{M}_{\mathrm{X}}^{4}+14 \mathcal{M}_{\mathrm{X}}^{2}-3)/16 \mathcal{M}_{\mathrm{X}}^{2},
\end{array}\right.
\end{equation}
where $i=1,2$ represents upstream and downstream regions, respectively.
The postshock magnetic field is then boosted as
\begin{equation}
B_{2} = B_{1}\sqrt{\frac{1}{3}+\frac{2\sigma^{2}}{3}},
\label{equ:post_Bfield}
\end{equation}
where $\sigma = \rho_{2}/\rho_{1}$ \citep{Kang2017}.

\subsection{Evolution of Merger Shocks}
\label{section:shock_model}

\begin{figure}
\includegraphics[width=\columnwidth]{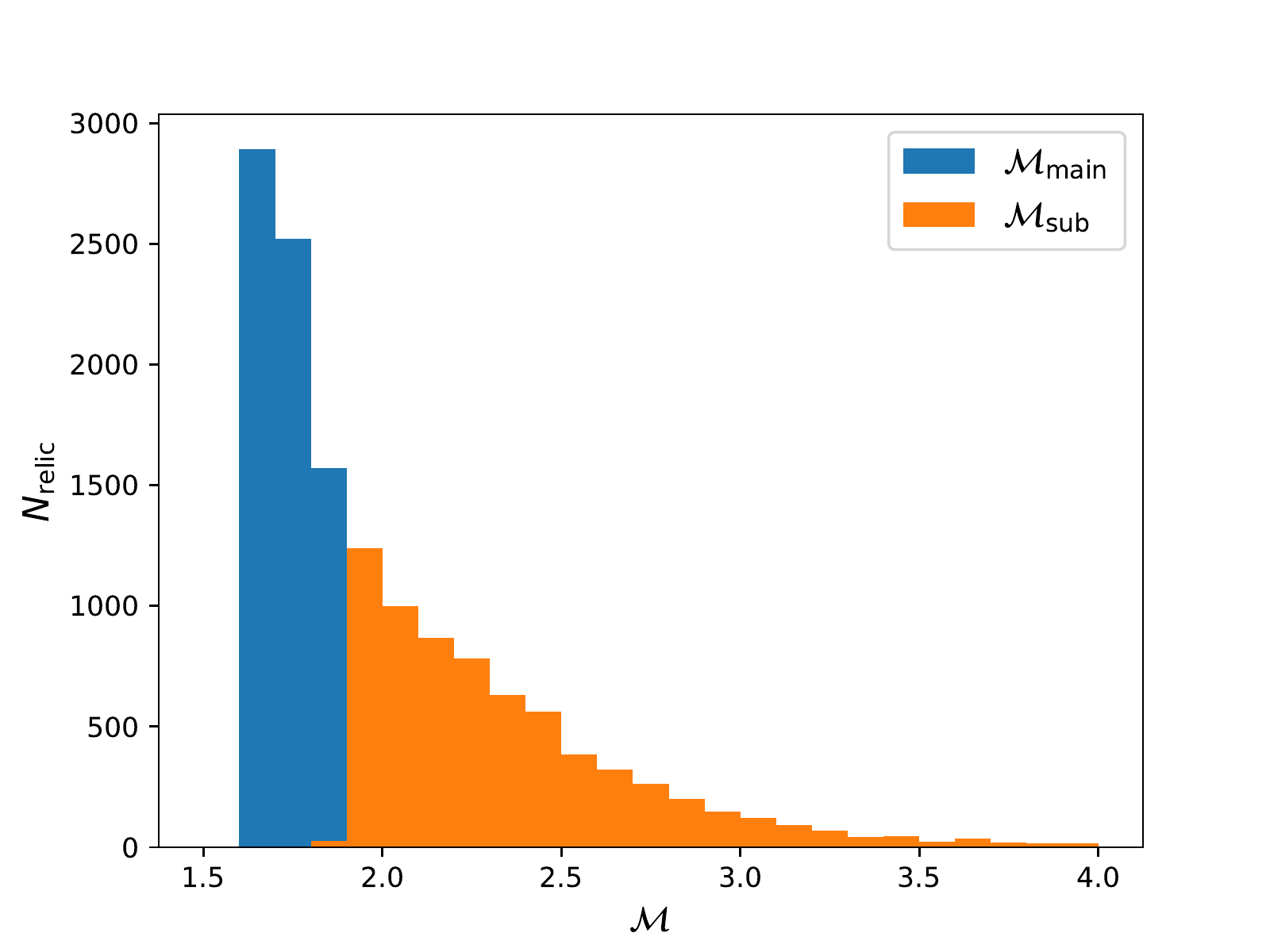}
\caption{Distribution of the Mach numbers of the simulated merger shocks. \label{fig:mach_dis}}
\end{figure}

\begin{figure}
\includegraphics[width=\columnwidth]{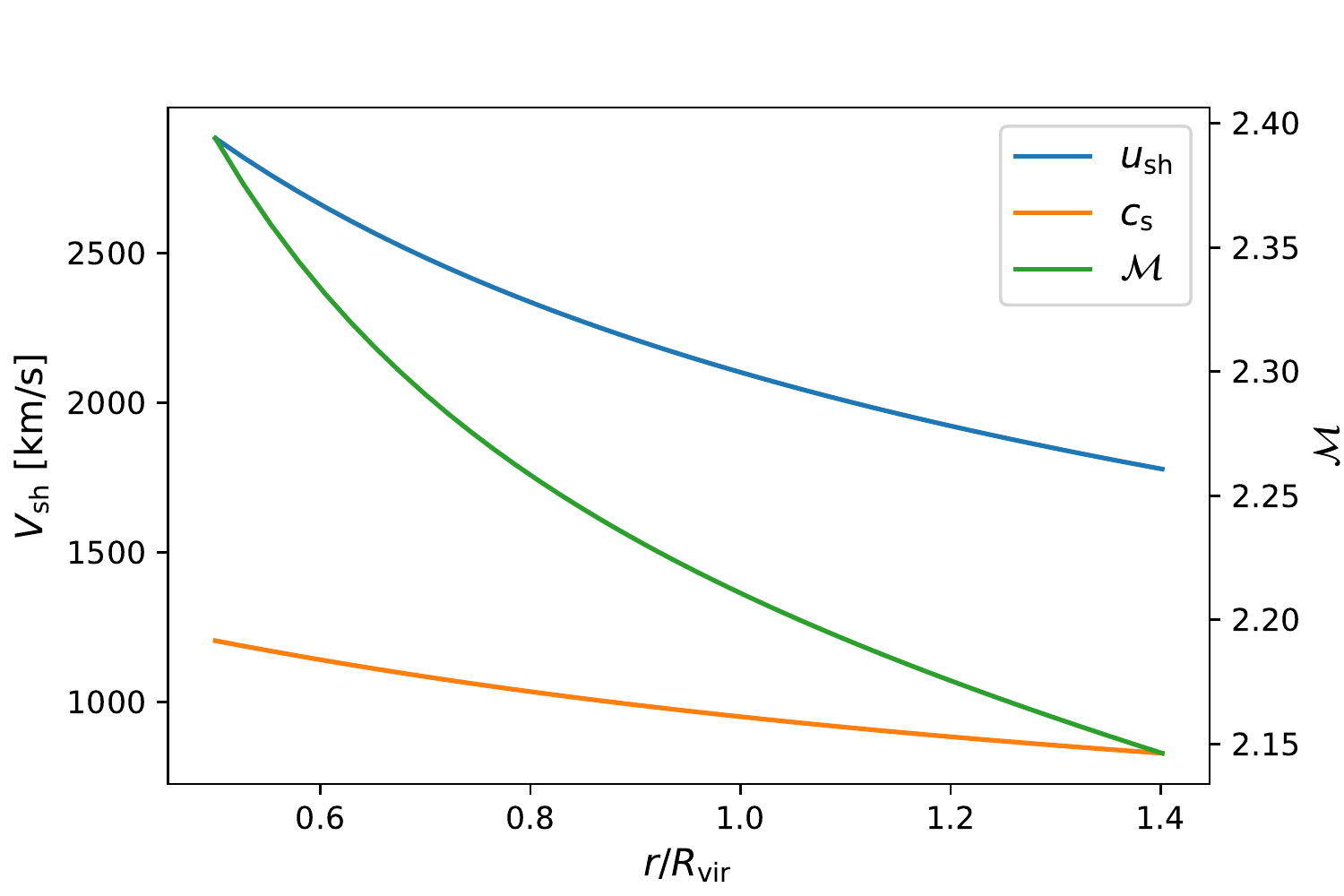}
\caption{Radial evolutions of the shock speed $u_{sh}$, sound speed $c_{s}$ and Mach number $\mathcal{M}$ for a binary merger with $M_{1} = M_{2} = 5\times 10^{14}\, \text{M}_{\odot}$ at $z=0.2$. \label{fig:shock_evol}}
\end{figure}

The properties of the merger shocks vary remarkably from case to case, depending on both merger parameters (e.g., mass ratio, initial velocity, and impact parameter) and evolution stage of the shock.
Although a few radio relics are detected in cool-core systems, which are possibly related to off-axis minor mergers (i.e., the mergers with non-zero impact parameter and mass ratio $\gtrsim3$; \citealt{Feretti}), observational evidences show that most radio relics are likely to appear in mergers with a small impact parameter \citep{Hoang2018},
so we limit our study to binary head-on mergers, i.e., we always set the impact parameter to be zero. Therefore, two shocks will be produced along the merger axis, for which the initial Mach numbers can be given in terms of the mass ratio between the subcluster and the main cluster $\eta_{\mathrm{m}}=M_{2}/M_{1}$ ($M_1>M_2$) \citep{Takizawa_1999, Gabici2003_RRmodel}
\begin{equation}\label{equ:shock_mach}
\left\{\begin{array}{l}
\mathcal{M}_{\mathrm{main}}^{2}=\frac{4(1+\eta_{\mathrm{m}})}{\gamma(\gamma-1)}\left[\frac{1}{1+\eta_{\mathrm{m}}^{1 / 3}}-\frac{1}{4(1+\eta_{\mathrm{m}})^{1 / 3}}\right],
\\
\\
\mathcal{M}_{\mathrm{sub}}^{2}=\eta_{\mathrm{m}}^{-2 / 3} \mathcal{M}_{\mathrm{main}}^{2},
\end{array}\right.
\end{equation}
where $\mathcal{M}_{\mathrm{main}}$ and $\mathcal{M}_{\mathrm{sub}}$ denote the Mach numbers of the shocks propagating in the main cluster and subcluster, respectively. \edited{The calculated Mach numbers are plotted in Figure~\ref{fig:mach_dis} for our simulated shocks.} 

On the other hand, because radio relics are rare in the central region of clusters, possibly due to the fact that 
the comoving kinetic power through the shock surface
is generally small in central regions \citep{Vazza2012_rare_centraRR}, we assume that the shocks start to accelerate electrons at $R_{\mathrm{sh,min}} = 0.5\,R_{\mathrm{vir}}$ \citep{Feretti}, \edited{and} the acceleration begins $\Delta t \sim 0.5\,R_{\mathrm{vir}}/u_{\mathrm{sh,0.5}}$ after the merger occurs, where \edited{$u_{\mathrm{sh,0.5}}=\mathcal{M}\cdot c_{\mathrm{s}}$} is the shock velocity at $r=0.5\,R_{\mathrm{vir}}$ \edited{with the mach number $\mathcal{M}$ given in equation~\ref{equ:shock_mach} and the sound speed $c_{\mathrm{s}} = 150\,\text{km}\,\text{s}^{-1}\left(T/10^{6}\text{K}\right)^{1/2}$ \citep{Kang_2010}, which is dependent on the ICM temperature $T$.}
To describe the evolution of the shock's strength, we apply the analytical model provided in \citep{Zhang2019} and assume that shock velocity $u_{\mathrm{sh}}$ measured in the rest frame of the upstream ICM decreases with shock's position $R_{\mathrm{sh}}$ following a power law form \edited{
\begin{equation}
u_{\mathrm{sh}}= u_{\mathrm{sh,0.5}} \left(\frac{R_{\mathrm{sh}}}{0.5R_{\mathrm{vir}}}\right)^{\eta},
\end{equation}
}where $\eta=\omega / 4 -1$ and $\omega = - d\ln{\rho}/d\ln{r}$  ($\simeq 2$ for the $\beta$ model). As an example,
in Figure~\ref{fig:shock_evol} we show the radial evolution of shock speed $u_{\mathrm{sh}}$, sound speed $c_{\mathrm{s}}$, and Mach number $\mathcal{M} = u_{\mathrm{sh}}/c_{\mathrm{s}}$ \edited{at $0.5R_{\mathrm{vir}}$} for a binary merger with $M_{1} = M_{2} = 5\times 10^{14}\,\text{M}_{\odot}$, and $z=0.2$. 
\edited{The plots in this section following (Figure \ref{fig:fp_atshock} and Figure \ref{fig:diff_R}) are all based on this example.}
We assume that during the propagation a shock stays as part of a spherical surface and has a constant solid angle $\Omega$ with the cluster center being the apex. Thus the surface area of the shock grows with the radius as $A \propto r^{2}$.

\subsection{Shock Acceleration}
\label{section:DSA_solution}
Within the frame of DSA theory, the behavior of electrons in both real space and momentum space is described by the diffusion-convection equation \citep{Skilling1975}
\begin{equation}\label{equ:diffusion-convection}
\begin{gathered}
\frac{\partial f(r,p)}{\partial t} + u\frac{\partial f(r,p)}{\partial r}=
\frac{1}{3} \frac{\partial}{\partial r}\left[up\frac{\partial f(r,p)}{\partial p}\right]\\
+ \frac{\partial}{\partial r}\left[\kappa(r, p) \frac{\partial f(r,p)}{\partial r}\right]
+\frac{1}{p^{2}} \frac{\partial}{\partial p}\left[p^{2} D_{pp}\frac{\partial f(r,p)}{\partial p}\right],
\end{gathered}
\end{equation}
where $f(r,p)$ is the number of electrons with momentum $p$ per unit volume, $u$ is the background flow velocity, $r$ is the distance to the cluster center, $\kappa(r,p)$ and $D_{pp}$ are the spatial and momentum diffusion coefficients, respectively. 
Note that the equation is expressed in the rest frame of the shock and the energy loss of relativistic electrons (e.g., via radio synchrotron) is not included. On the right side of the equation the first term describes the influence of the ICM's bulk flow, which corresponds to the convection process, 
the second term the diffusion of electrons in real space, and the third term the diffusion in the momentum space that is caused by the turbulence behind the shock. 
By applying the gas dynamic conservation equations and considering the feedback of CR on the shock and ICM, we may numerically solve equation~\ref{equ:diffusion-convection} \citep{Kang2012a, Kang2012b}, which is, however, usually computationally expensive. In our case only weak shocks are considered, therefore the CR feedback is insignificant and can be ignored in the calculation (i.e., the test-particle limit). Since the thermal and fossil relativistic electron populations are accelerated independently at the shock front, we describe how to calculate the number distributions of the electrons for the two populations in Section~\ref{section:DSA_thermal} and~\ref{section:DSA_fossil} separately. 

\subsubsection{Contribution of the Thermal Electron Population}
\label{section:DSA_thermal}
We follow the approach of, e.g., \citet{Hoeft2007_RRsimu} and \citet{Vazza2020}, and ignore the contribution of turbulence acceleration, which may spatially widen the relics to some extent but without adjusting the radio properties of the relics significantly \citep{Kang2017}.
For thermal electrons equation~\ref{equ:diffusion-convection} can be analytically solved after the convection and diffusion processes achieve a balance for a one-dimension shock front
\begin{equation}\label{equ:DSA_thermal_acc}
f(p)_{\mathrm{th,acc}} = f_{\text{N,th}}\left(\frac{p}{p_{\mathrm{inj}}}\right)^{-q} \exp \left(-\frac{p^{2}}{p_{\mathrm{eq}}^{2}}\right),
\end{equation}
where $f_{\text{N,th}}$ is the normalization factor, $q = 3\sigma/(\sigma-1)$ \citep{Drury_1983} with the density compression factor $\sigma=\rho_{2}/\rho_{1}$ given in equation~\ref{equ:RH}, and $p_{\mathrm{inj}}$ and $p_{\mathrm{eq}}$ are the injection momentum and cutoff momentum, respectively. 
Since the amount of the energy gained by the electrons from a single passage across the shock front is small \citep{Fermi_1949}, the electrons responsible for the radio relics must have recrossed the shock repeatedly, which requires that the gyroradii of the electrons in the magnetic field be at least several times of the shock's thickness. 
Since electrons with lower energies have smaller gyroradii, there should exist a threshold momentum $p_{\mathrm{inj}}$ for the electrons to participate in the DSA
\citep{Kang_2002}. 
According to hybrid simulation of \citet{Kang2017} and \citet{Caprioli_2014}, $p_{\mathrm{inj}}\sim 150\,p_{\mathrm{th,e}}$, where $p_{\mathrm{th,e}} = \sqrt{2m_{\mathrm{e}}kT_{2}}$. The cutoff momentum $p_{\mathrm{eq}}$, on the other hand, can be determined by letting the acceleration rate equal to the loss rate caused by synchrotron emission and inverse Compton scattering off the CMB photons \citep{Kang2011}
\begin{equation}\label{equ:p_eq}
p_{\mathrm{eq}}=\frac{m_{\mathrm{e}}^{2} c^{2} u_{\mathrm{sh}}}{\sqrt{4 e^{3} q / 27}}\left(\frac{B_{1}}{B_{\mathrm{e}, 1}^{2}+B_{\mathrm{e}, 2}^{2}}\right)^{1 / 2} k_{\mathrm{Bohm}}^{-1},
\end{equation}
where $i=1,2$ represents the upstream and downstream regions, respectively, $B_{\mathrm{e}}^{2} = B^{2} + B_{\mathrm{rad}}^{2}$ is the total effective magnetic field ($B$ is the magnetic field of ICM and $B_{\mathrm{rad}}^{2}/8\pi$ is the energy density of ambient radiation field \cite{longair_1994}), and $k_{\mathrm{Bohm}}\sim 1$ corresponds to Bohm diffusion coefficient \citep{Kang2017}.

We determine $f_{\text{N,th}}$ in equation~\ref{equ:DSA_thermal_acc} by assuming that the energy spectrum of the electrons is continuous at $p_{\mathrm{inj}}$, i.e. $f_{\mathrm{th}}(p_{\mathrm{inj}}) = f_{\mathrm{th,acc}}(p_{\mathrm{inj}})$, where $f_{\mathrm{th}}$ and $f_{\mathrm{th,acc}}$ are the number density distributions of the thermal electrons before and after being accelerated, respectively. 
Meanwhile, because additional pre-acceleration processes such as shock drift acceleration (SDA) and electrons firehose instability (EFI) \citep{Kang_2019_preacc, Trotta2019_preacc} that may occur ahead of the shock tend to change the initial energy distribution of the thermal electrons, which can cause a large high energy tail in the energy spectrum, we use a $\kappa$ distribution instead of the Maxwell distribution to present the thermal electron population \citep{Kang2017}
\edited{
\begin{equation}\label{equ:kappa_dist}
f(p)_{\kappa} \propto (1+\frac{p^{2}}{\kappa\,p_{\mathrm{th,e}}^{2}})^{-(\kappa +1)} \exp \left(-\frac{p^{2}}{p_{\mathrm{inj}}^{2}}\right),
\end{equation}}
where $\kappa$ is a constant.
\edited{The cut-off term $ \exp \left(-p^{2}/p_{\mathrm{inj}}^{2}\right)$ is included to make sure the thermal population could hardly has electrons with $p>p_{\mathrm{inj}}$ since when the electrons in front of the shock are accelerated to the $p_{\mathrm{inj}}$, most of them would join the DSA process and then would not reflect back to the preshock region.}
SDA and EFI also happen in the solar winds, the CRe (CR electron) spectrum of which is usually well fitted with a $\kappa$ distribution with $2<\kappa<5$ \citep{Pierrard2010}, and we choose to take $\kappa =2$ because $\kappa >2$ the resulting radio relics are generally too dim.

\begin{figure}
\includegraphics[width=\columnwidth]{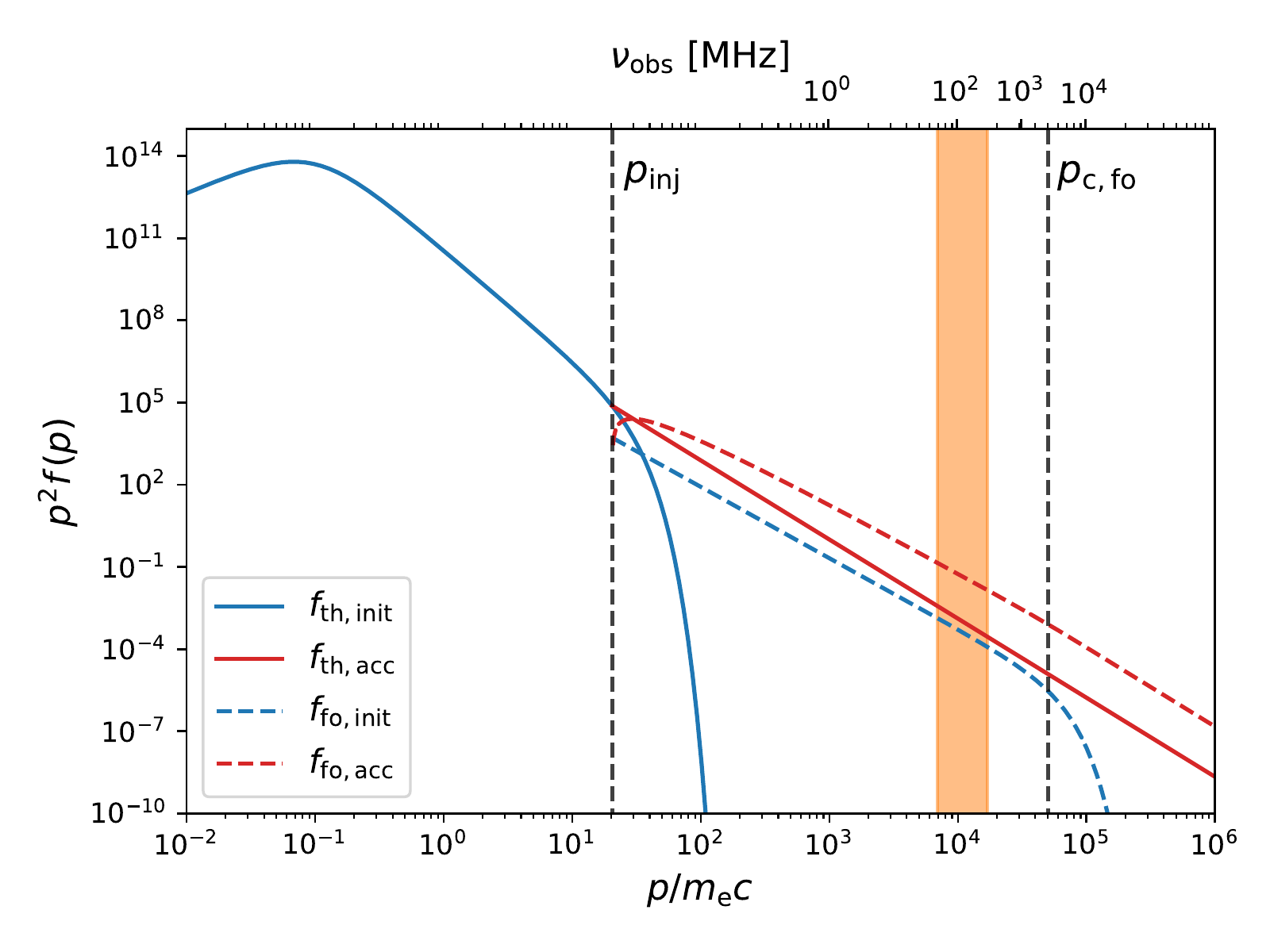}
\caption{$p^{2}f(p)$ distributions of the thermal electron population (blue) and the fossil relativistic electron population (red) before and immediately after shock acceleration (solid and dash, respectively).
The vertical black dash lines indicate $p_{\mathrm{inj}}$ and $p_{\mathrm{c,fo}}$.
X-axis: electron energy and corresponding characteristic frequency of synchrotron emission. 
\edited{
The shaded area marks 50 $\sim$ 350 MHz, which is the target band of the SKA1-Low.}
\label{fig:fp_atshock}}
\end{figure}

\subsubsection{Contribution of Fossil Electrons Population}
\label{section:DSA_fossil}
Following \citet{Kang2017} we assume that the fossil electrons initially present a power-law spectrum with an exponential cutoff at $p_{\mathrm{c,fo}}$
\begin{equation}\label{equ:fossil_init}
f_{\mathrm{fo,init}}(p)=f_{\text{N,fo}} \cdot p^{-s} \exp \left[-\left(\frac{p}{p_{\mathrm{c,fo}}}\right)^{2}\right],
\end{equation}
where $f_{\text{N,fo}}$ is the normalization, $s$ is the initial energy spectral index, and the cutoff momentum $p_{\mathrm{c,fo}} = 5 \times 10^{4}\ m_{\mathrm{e}}c$, \edited{which could reproduce the observed flux density $S_{\nu}$ and spectral index $\alpha$ of CIZA J2242.8+5301 \citep{Kang2016} and RX J0603.3+4212 \citep{Kang2017}.
According to our test, the resulting simulated power of relics is insensitive to the value of $p_{\mathrm{inj}}$.}
Under the same conditions applied to obtain equation~\ref{equ:DSA_thermal_acc}, i.e., the effect of turbulence is neglected and a balance between convention and diffusion processes has been established, a solution can be found in a integral form for the shock reaccelerated fossil electrons \citep{Drury_1983}
\begin{equation}\label{equ:DSA_solution_fossil}
f_{\text {fo,acc}}\left( p\right)=q \cdot p^{-q} \int_{p_{\text {inj }}}^{p} p^{\prime q-1} f_{\text{fo,init}}\left(p^{\prime}\right) d p^{\prime}.
\end{equation}

We show the derived energy spectra \edited{at $0.5R_{\mathrm{vir}}$} of the two electron populations in Figure~\ref{fig:fp_atshock}, \edited{whose magnetic field strength is given by equation \ref{equ:B_profile} and equation \ref{equ:post_Bfield}, and $s=4.6$, $f(p_{\mathrm{inj}})_{\mathrm{fo,init}}/f(p_{\mathrm{inj}})_{\mathrm{th,init}}=0.86447$}.
We also plot the characteristic frequency ($v_{\mathrm{obs}}$) of the synchrotron radiation corresponding to a given electron momentum using the relation \citep{Kang2012b}
\begin{equation}\label{equ:syn_gamma_nu}
\gamma \approx 1.26 \times 10^{4}\left(\frac{v_{\mathrm{obs}}}{1 \mathrm{GHz}}\right)^{\frac{1} {2}}\left(\frac{B}{5\,\mu \mathrm{G}}\right)^{-\frac{1}{2}}(1+z)^{\frac{1}{2}},
\end{equation}
and the band of the SKA1-low array (50 $\sim$ \edited{350 MHz}).

\subsection{Evolution of Electron Energy Spectra}
\label{section:spe_evo}
\edited{Besides computing the spectra of the accelerated electrons injected at the shock front, we also consider the influence of the radiation cooling, i.e., the aged CRe population downstream.}
We have assumed that both thermal and fossil electrons are accelerated instantaneously at shock front since the acceleration timescale is much shorter than the decay timescale ( $\lesssim$ 0.1 Gyr; \citealt{Brunetti2014_CReReview}). Thus we may apply the solutions of equations~\ref{equ:DSA_thermal_acc} and \ref{equ:DSA_solution_fossil} as the initial conditions to investigate the energy loss rate of relativistic electrons $(d\gamma/dt)$ behind the shock, which is attributed to three major processes, i.e., radio synchrotron radiation, inverse Compton scattering (IC) and Coulomb collision. In the first two processes the energy loss rates are \citep{Sarazin1999}
\begin{equation}\label{equ:syn_loss_rate}
\left(\frac{d \gamma}{d t}\right)_{\mathrm{syn}}=-4.10 \times 10^{-5} \gamma^{2}\left(\frac{B_2}{1\,\mu \mathrm{G}}\right)^{2}\quad\left[\mathrm{Gyr}^{-1}\right],
\end{equation}
where $B_{2}$ is the magnetic field in the postshock region, and 
\begin{equation}\label{equ:IC_loss_rate}
\left(\frac{d \gamma}{d t}\right)_{\mathrm{IC}}=-4.32 \times 10^{-4} \gamma^{2}(1+z)^{4} \quad\left[\mathrm{Gyr}^{-1}\right],
\end{equation}
respectively, showing that the two processes have comparable contributions and dominate the energy loss when the electrons energy is high ($\gamma \gtrsim 200$). 
In the case of Coulomb collision, the energy loss rate is given by \citep{Sarazin1999}
\begin{equation}\label{equ:CC_loss_rate}
\begin{gathered}
\left(\frac{d \gamma}{d t}\right)_{\mathrm{Coul}}=-3.79 \times 10^{4}\left(\frac{n_{\mathrm{th}}}{1 \mathrm{~cm}^{-3}}\right)\\ \times\left[1+\frac{1}{75} \ln \left(\gamma \frac{1 \mathrm{~cm}^{-3}}{n_{\mathrm{th}}}\right)\right] \quad\left[\mathrm{Gyr}^{-1}\right] .
\end{gathered}
\end{equation}
Since the energy loss rates due to synchrotron emission and inverse Compton scattering both have the form of  $d\gamma/dt = b\gamma^{2}$ in a steady environment, where $b$ is a constant, the corresponding electron energy spectra will have an analytical solution at any time if the effect of the Coulomb collision is negligible; otherwise the equation needs to be solved numerically. Considering that the impact of Coulomb collision is usually limited to $\gamma \lesssim 10^{3}$ (see Figure~\ref{fig:loss_rate} in Section~\ref{section:CC}) with which the electrons can barely produce detectable radio emission, it is reasonable to ignore its contribution in our model. The energy spectrum at any time $t$ after the shock acceleration is then given by \citep{Wang_2010}
\begin{equation}
f\left(\gamma, t\right) = \frac{f\left(\gamma_{0}, t_{0}\right)}{\left(1-b\gamma t\right)^{2}},
\end{equation}
where $\gamma_{0}=\gamma/(1-b\gamma t)$ is the initial electron energy.

\begin{figure}
\includegraphics[width=\columnwidth]{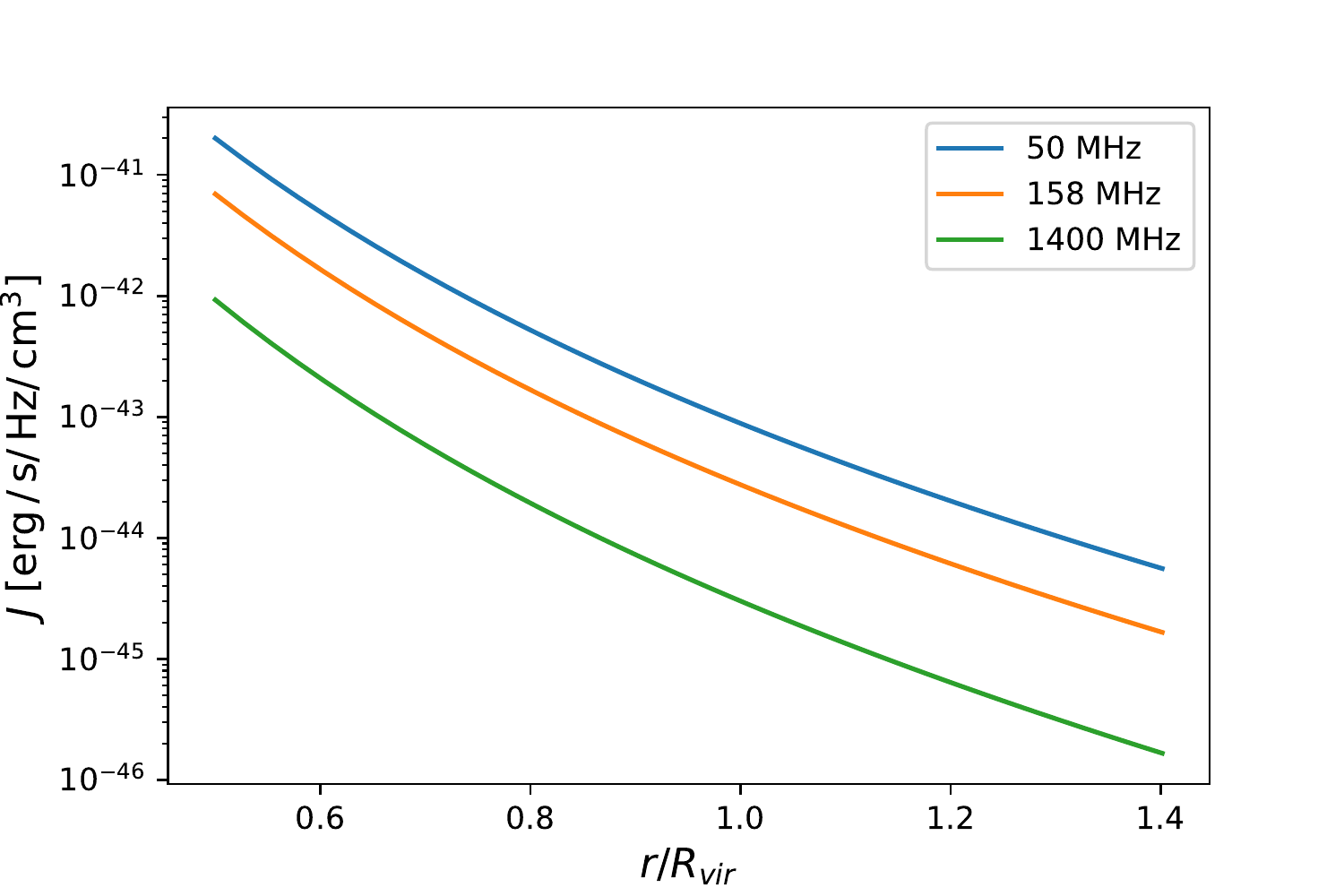}
\caption{The radio synchrotron emissivity of different radii immediately after shock fronts
\label{fig:diff_R}}
\end{figure}

\subsection{Radio Synchrotron Emission of Radio Relics}
\label{section:radioemi}
Given the number density distribution $f(\gamma,t)$, the synchrotron emissivity of the shock accelerated electrons in the radio relic is given by
\begin{equation}\label{equ:syn_emit}
J(\nu)=\frac{\sqrt{3} e^{3} B}{m_{\mathrm{e}} c^{2}} \int_{\gamma_{\min}}^{\gamma_{\max}} \int_{0}^{\frac{\pi}{2}} F_{\mathrm{syn}}\left(\frac{\nu}{\nu_{\mathrm{c}}}\right) f(\gamma, t) \sin ^{2} \theta \mathrm{d} \theta \mathrm{d} \gamma,
\end{equation}
where $\theta$ is the pitch angle of electrons with respect to the magnetic field, $\nu_{\text{c}} = 3\,\gamma^{2}\,\nu_{\text{L}}\,\text{sin}\theta/2$ is the critical frequency for a given Larmor frequency $\nu_{\text{L}} = eB/(2\pi m_{\text{e}}c)$, and $F(x)$ is the synchrotron kernel
\begin{equation}\label{equ:syn_kernel}
F_{\text {syn }}(x)=x \int_{x}^{\infty} K_{5 / 3}(y) \mathrm{d} y,
\end{equation}
where $K_{5/3}(y)$ is the modified Bessel function of order 5/3 \citep{Rybicki1979}.

Because we assume that shock acceleration starts to occur at $0.5\,R_{\mathrm{vir}}$ and the solid angle of the shock front always stays the same during the propagation, the volume element swept by the shock during $dt$ is $dV = \Omega\,R_{\mathrm{sh}}^{2}\,  dR_{\mathrm{sh}} = \Omega\,R_{\mathrm{sh}}^{2} u_{\mathrm{sh}}dt$. Although the shock may travel a long distance in a cluster's outskirts \citep{Zhang2019}, it is not clear to what extend can effective electron acceleration and radio synchrotron emission persist given the lower gas density and magnetic field. We calculate the spatial evolution of shocks in the range of $0.25\,R_{\mathrm{vir}} \sim 1.7\,R_{\mathrm{vir}}$, and find that at $1.2\,R_{\mathrm{vir}}$ the radio emission is only about $0.1\%$ of that produced at $0.5\,R_{\mathrm{vir}}$, \edited{which is shown in the Figure \ref{fig:diff_R}}. Thus it is reasonable to set a  maximum radius $R_{\mathrm{sh,max}}=1.2\,R_{\mathrm{vir}}$ (i.e., $0.5\,R_{\mathrm{vir}}<R_{\mathrm{sh}}<1.2\,R_{\mathrm{vir}}$).
When a shock travels to $R_{\mathrm{sh}}>1.2\,R_{\mathrm{vir}}$, only radiative cooling is taken into account in the calculation since the shock acceleration is quenched. 
Also we define the evolution time of radio relics as $t_{\mathrm{evol}} = t_{\mathrm{obs}} - t_{\mathrm{merge}}$, where $t_{\mathrm{merger}}$ is the cosmological age when the merger begins to occur. 
We assume that $\Omega$ does not vary with frequency drastically since $\Omega$ is mainly decided by the spatial distribution of the accelerated electrons right behind the shock front.

\section{Calculation and Results}
\label{section:results}

\subsection{Observational constrains}
\label{section:real_case}
In this section we apply the observational constrains obtained from the radio relics sample of \citet{Feretti} and that in Table~\ref{table:relics_samples} on our model to determine the solid angle $\Omega$ of the shock, the normalization $f_{\text{N,fo}}$ and the initial energy spectral index $s$ of the fossil electrons (equation~\ref{equ:fossil_init}).

\subsubsection{Solid Angles of Radio Relics}
\label{section:solid_angle}
\begin{figure}
\includegraphics[width=0.75\columnwidth]{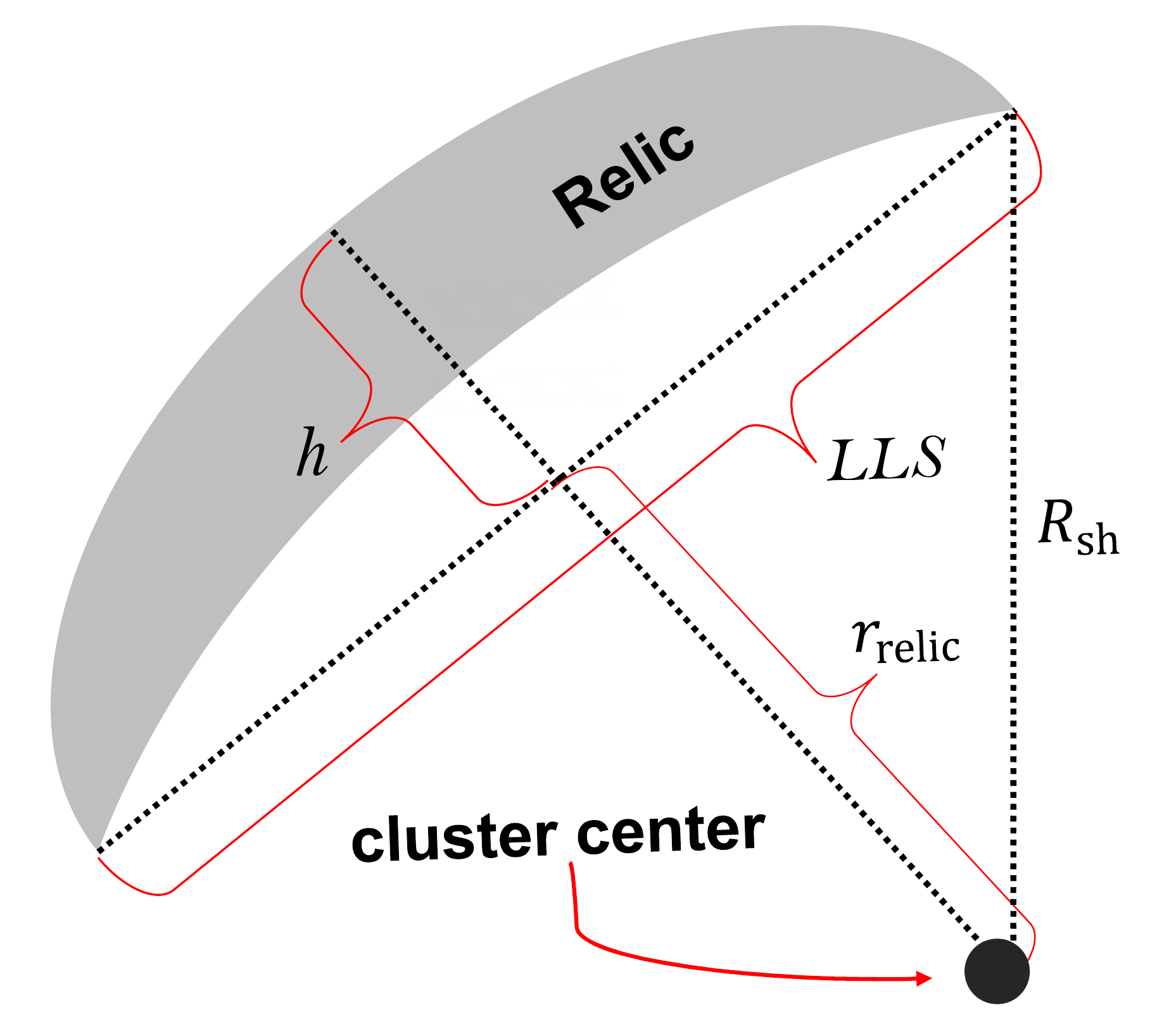}
\caption{\edited{The geometry model for a typical relic example.} \label{fig:geo_model}}
\end{figure}

\begin{figure}
\includegraphics[width=\columnwidth]{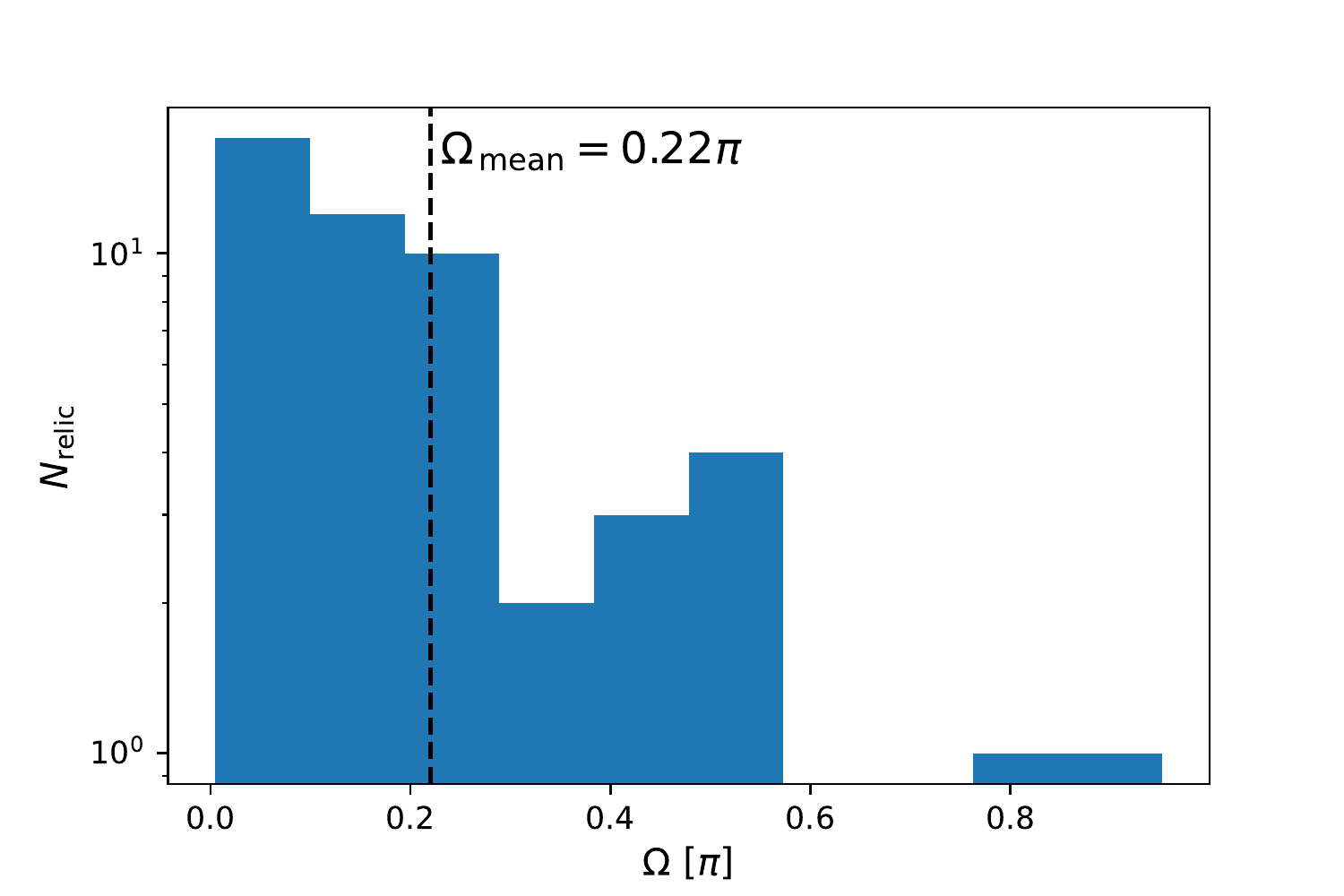}
\caption{Distribution of radio relics' solid angle $\Omega$ obtained from the sample of \citet{Feretti}. The dash line indicates the average value ($\Omega_{\mathrm{mean}}=0.22\pi$). \label{fig:omega_dis}}
\end{figure}

We use the data of 50 radio relics analyzed in \citet{Feretti} who listed the LLS and the distance to the cluster center $r_{\mathrm{relic}}$ for each radio relic, both observed at 1.4 GHz, to constrain $\Omega$. \edited{We treat radio relics as spherical crowns since we only study the binary head-on mergers in this work, which are symmetric along the merger axes, and we illustrate the geometry model in Figure \ref{fig:geo_model}.}
We calculate the solid angle as $\Omega = 2\pi h/R_{\mathrm{sh}}$, where $R_{\mathrm{sh}} = \sqrt{r_{\mathrm{relic}}^{2}+(\text{LLS}/2)^{2}}$ and \edited{$h = R_{\mathrm{sh}}-r_{\mathrm{relic}}$} is the height of the spherical crown. 
Based on the results (Figure~\ref{fig:omega_dis}) we randomly choose $\Omega$ for each simulated shock. 

\subsubsection{Fossil Electron Properties}
\label{section:case_fossil}

\begin{figure}
\includegraphics[width=\columnwidth]{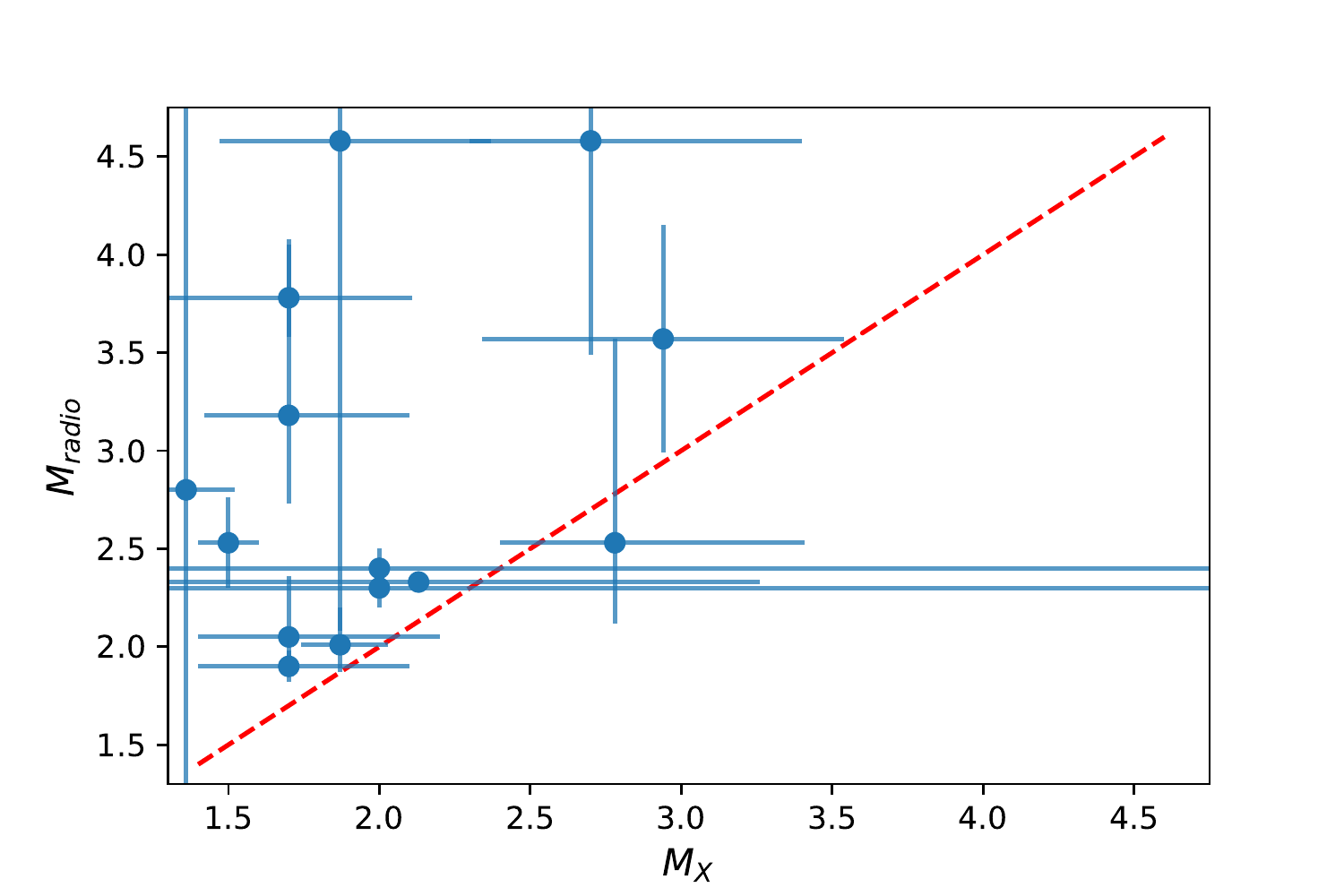}
\caption{Distribution of $\mathcal{M}_{\mathrm{X}}$ and $\mathcal{M}_{\mathrm{radio}}$ of our radio relic sample. The red dash line indicates where $\mathcal{M}_{\mathrm{X}}$ = $\mathcal{M}_{\mathrm{radio}}$. \label{fig:machlist}}
\end{figure}

Although in the calculation of $\mathcal{M}_{\mathrm{radio}}$ the fossil electrons population is not included, making $\mathcal{M}_{\mathrm{radio}}$ inappropriate for representing the velocity of the shock, it can be used to calculate the energy spectral index of the relativistic electrons. Considering the synchrotron emissivity $J_{\mathrm{syn}}(\nu) \propto \nu^{-\alpha}$ and the spectral index \edited{at the shock front $\alpha_{\mathrm{inj}}=(\delta_{\mathrm{inj}}-3)/2$, where $\delta_{\mathrm{inj}}$ is the power law index of the electrons' energy distribution in three-dimensional momentum space}, the radio Mach number is given by \edited{$\mathcal{M}_{\mathrm{radio}} = \sqrt{1+4/(2\alpha_{\mathrm{inj}}-1)} = \sqrt{1+4/(\delta_{\mathrm{inj}}-4)}$}.

When only thermal electrons are considered (equation~\ref{equ:DSA_thermal_acc}), 
$\delta_{\mathrm{inj}} = q$, in which case $\mathcal{M}_{\mathrm{X}}=\mathcal{M}_{\mathrm{radio}}$. 
For fossil electrons,
if we ignore the exponent cutoff in equation~\ref{equ:fossil_init}, which only plays a role at $p\sim p_{\mathrm{inj}}$, equation~\ref{equ:DSA_solution_fossil} can be solved analytically
\edited{
\begin{equation}\label{equ:weak_shock_model}
f_{\mathrm{fo,acc}}(p)=\left\{\begin{array}{l}
\frac{q}{(q-s)} f_{\text{N,fo}}\left[\left(\frac{p}{p_{\text{inj}}}\right)^{-s}-\left(\frac{p}{p_{\text{inj}}}\right)^{-q}\right],
\\
\hfill \text { if } q \neq s ; \\
\\
q  f_{\text{N,fo}} \frac{p^{-q}}{p_{\text{inj}}^{s}} \ln \frac{p}{p_\text{inj}},
\hfill \text { if } q=s .\\
\end{array}\right.
\end{equation}
}
Therefore, for the fossil electron population with \edited{a flatter initial energy spectrum than that of the accelerated thermal electrons}, i.e., $s<q$, the dominant spectral index of resulting $f_{\mathrm{fo,acc}}$ is $s$, and $\mathcal{M}_{\mathrm{radio}} = \sqrt{1+4/(s-4)}$, which provides a possible explanation for $\mathcal{M}_{\mathrm{radio}} > \mathcal{M}_{\mathrm{X}}$. Adopting 
radio Mach number provided in Table~\ref{table:shock_samples}, we can calculate $s$ ($= 4\mathcal{M}_{\mathrm{radio}}^{2}/(\mathcal{M}_{\mathrm{radio}}^{2}-1)$) \edited{and further determine the shape of energy spectra of the potential fossil electrons in these observed samples}. 
We have to point out among our sample, there is one cluster: ACT-CL J0102-4915 with $\mathcal{M}_{\mathrm{X}}>\mathcal{M}_{\mathrm{radio}}$, which cannot be explained by equation~\ref{equ:weak_shock_model}. Considering the large observational error, we still include it in calculation to constrain the parameters.

\begin{figure}
\includegraphics[width=\columnwidth]{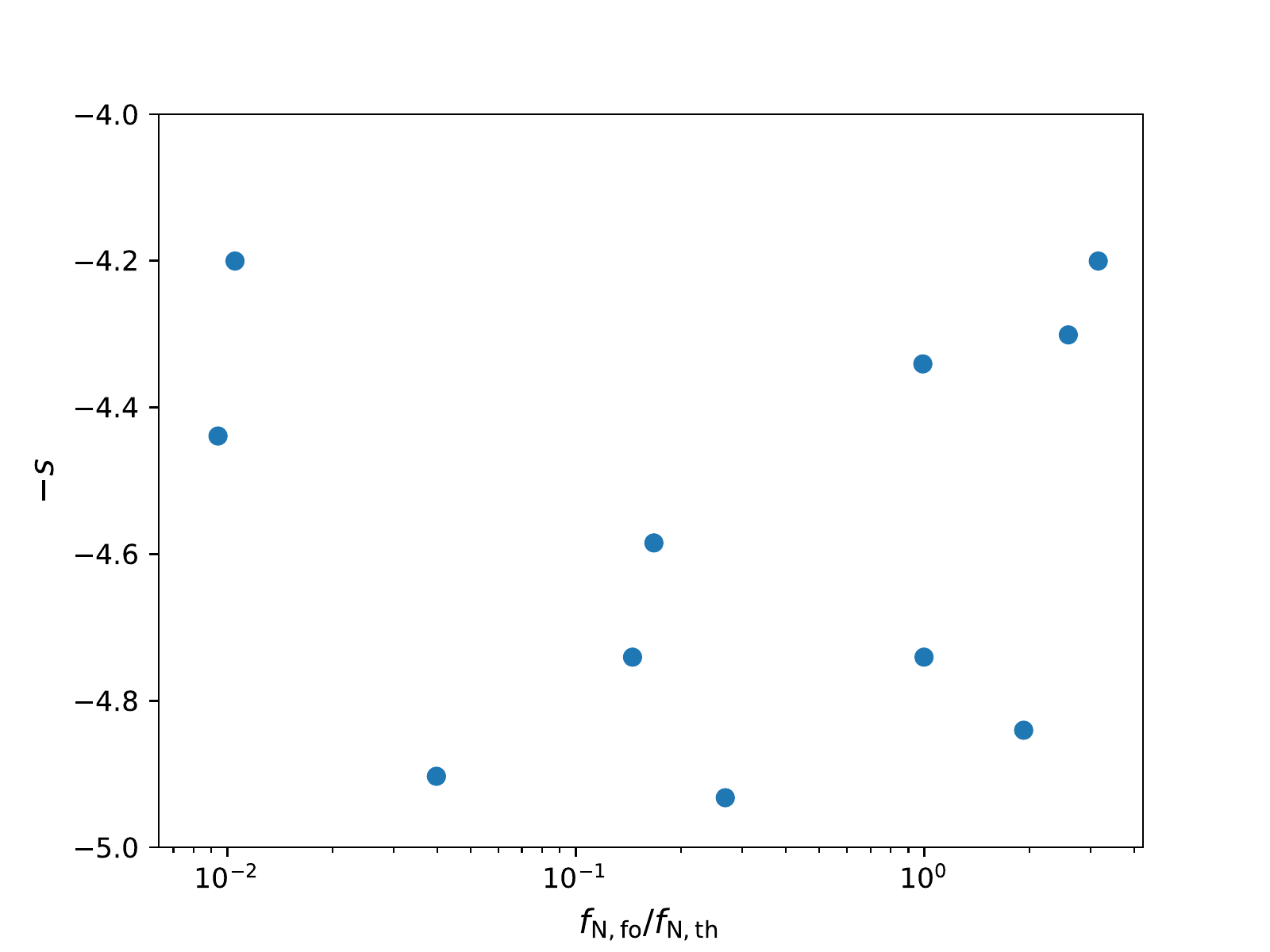}
\caption{Distribution of $-s$ against $f_{\mathrm{N,fo}}/f_{\mathrm{N,th}}$ calculated based on our model for the radio relics sample listed in Table~\ref{table:relics_samples}.}\label{fig:fossil_ratio}
\end{figure}

\edited{
Since the emission generated by fossil electrons ($P_{\mathrm{fossil}}$) is proportional to the particle density: $P_{\mathrm{fossil}}\propto f_{\mathrm{fo,acc}}(p)\propto f_{\mathrm{N,fo}}$ for a given cluster,  with the energy spectra of fossil electrons whose shape has been determined by the $\mathcal{M}_{\mathrm{radio}}$, we could calculate $P_{\mathrm{fo, unnorm}}$ based on our model (equation~\ref{equ:DSA_solution_fossil}), which equals to $P_{\mathrm{fossil}}/f_{\mathrm{N,fo}}$.
Then the normalized factor $f_{\mathrm{N,fo}}$ could be determined by
\begin{equation}
    f_{\mathrm{N,fo}} = \frac{P_{\text{fossil}}}{P_{\text{fo,unnorm}}} = \frac{P_{\text{observed}} - P_{\text{thermal}}}{P_{\text{fo,unnorm}}},
\end{equation}
where the $P_{\mathrm{observed}}$ is the observed power listed in Table~\ref{table:relics_samples}, and $P_{\mathrm{thermal}}$ is that produced by the thermal electrons, which could be computed with the equation~\ref{equ:DSA_thermal_acc}$\sim$\ref{equ:kappa_dist}.
}
We also calculate \edited{$\mathcal{P}_{\mathrm{CRe}}/\mathcal{P}_{\mathrm{total}}$}, the pressure ratio between the fossil electrons and thermal electrons, and list them in Table~\ref{table:relics_samples} along with $f_{\mathrm{N,fo}}/f_{\mathrm{N,th}}$.
We show the distribution of $-s$ and $f_{\mathrm{N,fo}}/f_{\mathrm{N,th}}$, the two parameters needed in our model, in Figure~\ref{fig:fossil_ratio}. In the simulation of radio relics, we apply \edited{$f_{\text{N,fo}} =  0.86447 f_{\text{N,th}}$}, which is the average value of our observed radio relics sample, and randomly choose $s$ between 4 and 5. 
\edited{It is worth noting that the $f_{\mathrm{N,fo}}/f_{\mathrm{N,th}}$ does not present the number density ratio between the two electron population since there are much more thermal electrons at $p< p_{\mathrm{inj}}$.}

\begin{figure}
\includegraphics[width=\columnwidth]{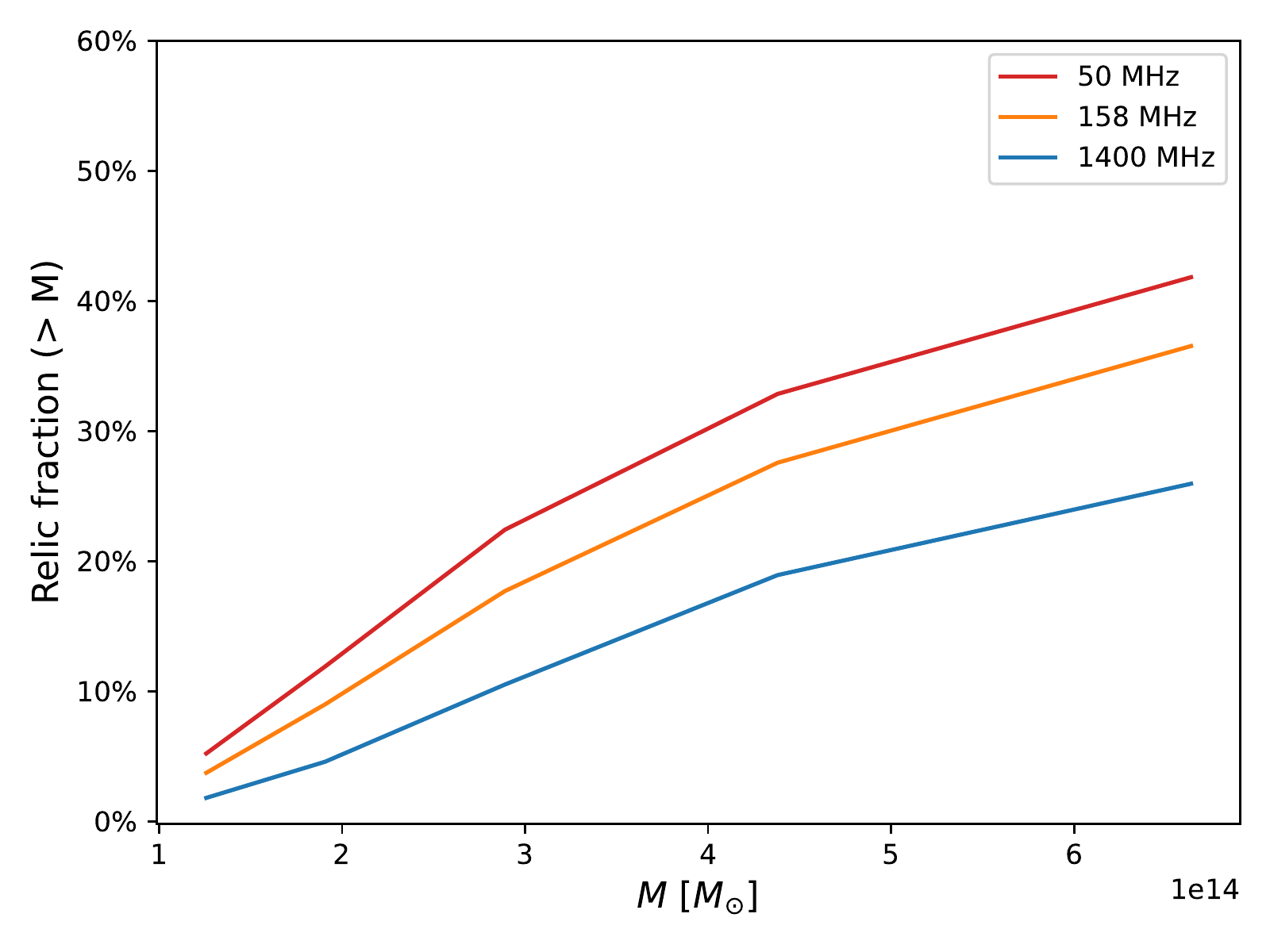}
\caption{\edited{Cumulative fraction of clusters hosting one or two radio relics whose flux larger than 1 $\mu$Jy as a function of the virial mass.}
\label{fig:Relic_frac}}
\end{figure}

\begin{figure}
\includegraphics[width=\columnwidth]{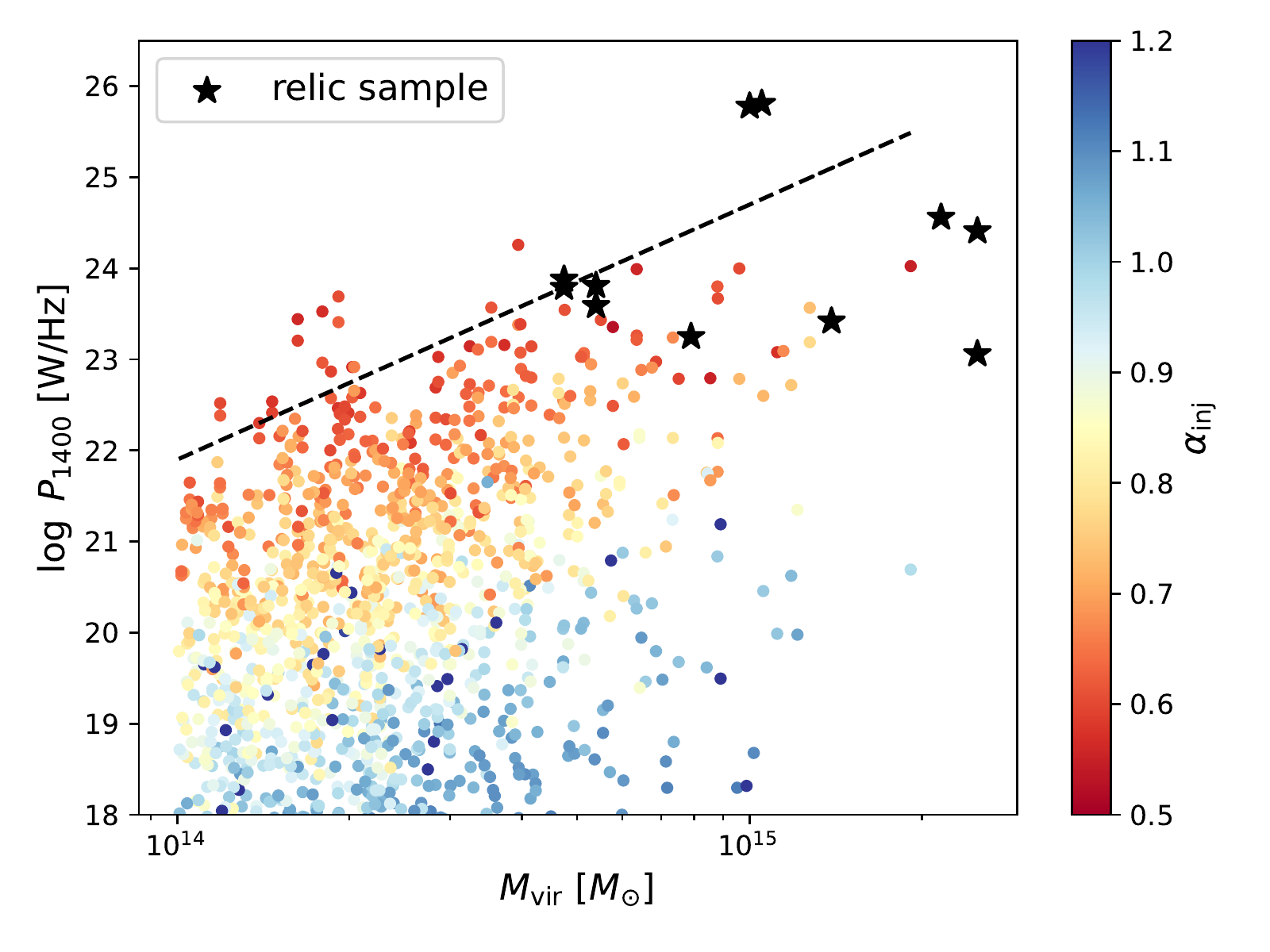}
\caption{1.4 GHz radio power versus virial mass for the simulated radio relics (dots; $t_{\mathrm{evol}}$ are marked with different colors). Black stars: observed radio relics listed in Table~\ref{table:relics_samples}. Black line: the $P_{\mathrm{1400}} \propto M_{\mathrm{vir}}^{\mathrm{2.8}}$ relation presented in \citet{DeGasperin2014_PM_scale}.
}\label{fig:P_1400_M}
\end{figure}

\begin{figure*}
\includegraphics[width=\columnwidth]{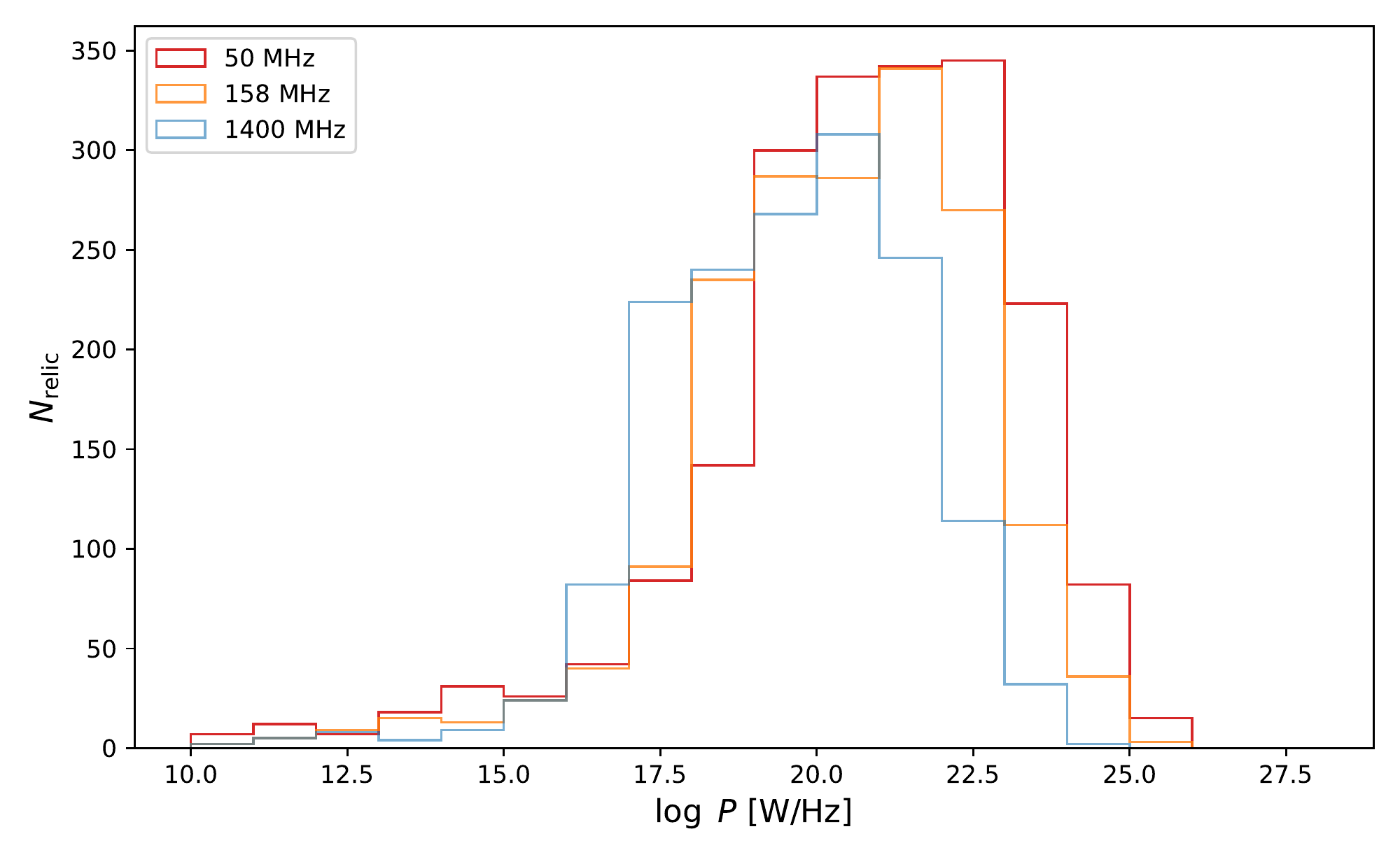}
\includegraphics[width=\columnwidth]{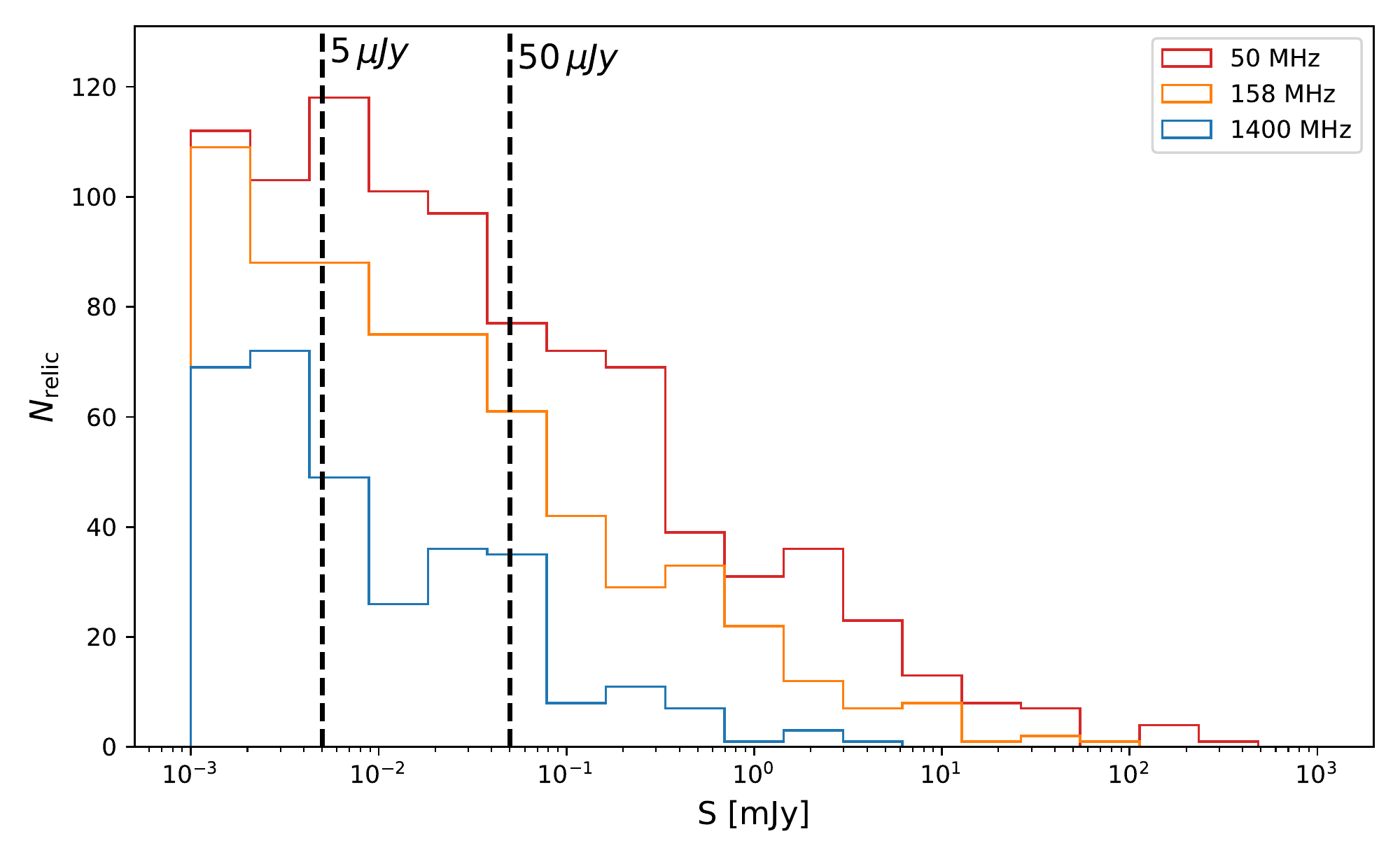}
\caption{ Distributions of radio powers (a) and fluxes (b) of simulated relics at 50 MHz, 158 MHz, 1.4 GHz. The dash lines ($S = 5\,\mu$Jy and $50\,\mu$Jy) mark the SKA and HERA sensitivities (100-hour observation), respectively.
}\label{fig:Results_P_S_dis}
\end{figure*}

\begin{figure}
\includegraphics[width=\columnwidth]{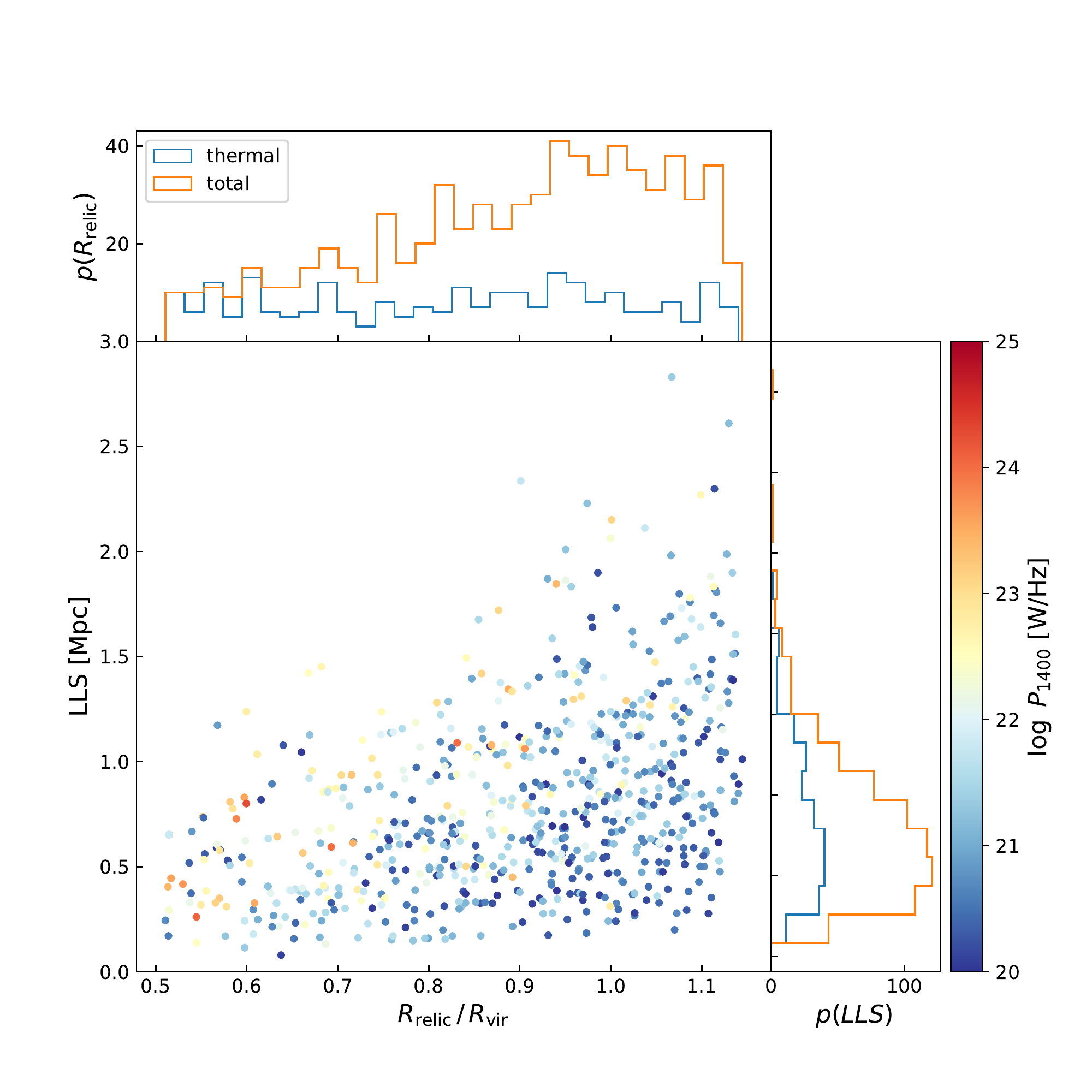}
\caption{Distribution of LLS and the $R_{\mathrm{relic}}$ of the simulated radio relics produced by total electrons (thermal electrons and fossil electrons) and solely by thermal electrons, respectively, with $P_{1400} > 10^{20}$ W/Hz. 
\label{fig:geo_dis}}
\end{figure}

\subsection{Simulated Radio Relics}
\label{section:simulated_cata}
By assuming that fossil electron population exists in 50\% of the galaxy clusters, which matches the fact that about 30\% of the massive clusters ($M_{\mathrm{vir}} > 5\times 10^{14}\,\text{M}_{\odot}$) host at least one radio relic \citep{Gleser2008_foreground} \edited{at 158 MHz},
which is shown in Figure~\ref{fig:Relic_frac}. \edited{Our model predicts that $9.6\%$ and $7.1\%$ clusters with $M_{\mathrm{vir}} > 1.2\times 10^{14}\,\mathrm{M}_{\odot}$ have one or more relics with $S>1\mu$Jy at 50 MHz and 158 MHz, respectively, which is consistent with the result of $10 \pm 6\%$ given by the Second Data Release of the LOFAR Two-meter Sky Survey (LoTSS DR2), whose target band is 120-168 MHz.}

In Figure~\ref{fig:P_1400_M} we show the $P_{1400}-M_{\mathrm{vir}}$ relation for both the simulated radio relics, \edited{which are colored by the injection radio spectral indices $\alpha_{\mathrm{inj}}$}, and the relics included in the observation sample (Table~\ref{table:relics_samples}). This shows that \edited{the upper limits of the power for a given $M_{\mathrm{vir}}$ in} our results agree very well with the relation $P_{1400} \propto M_{\mathrm{vir}}^{2.8}$ obtained by \citet{DeGasperin2014_PM_scale} based on the observations of 15 clusters. For clusters with \edited{smaller luminosity, there is a considerable scattering} 
and a tendency of deviation from $P_{1400} \propto M_{\mathrm{vir}}^{2.8}$,
possibly caused by  Malmquist bias \citep{Vazza2020}, i.e., only brightest radio relics are observed and included in the analysis in \citet{DeGasperin2014_PM_scale}. 
Along with Figure~\ref{fig:Results_P_S_dis}, this indicates that much more radio relics are to be found once the detection sensitivity is increased. 
\edited{The $\alpha_{\mathrm{inj}}$ are calculated by fitting straight power-laws on the simulation results at shock fronts at 50 MHz, 158 MHz and 1400 MHz.}

\edited{We find that out of 5107 merging clusters simulated in the $20^{\circ}\times 20^{\circ}$ sky patch, 1137, 1018 and 907 clusters host one or two radio relics brighter than $10^{15}$ W/Hz} at 50 MHz and 158 MHz, the characteristic frequencies that will be covered by the next-generation radio arrays, as well as 1.4 GHz, respectively (Figure~\ref{fig:Results_P_S_dis}). 
\edited{
The drop in the power distribution around $P\sim 10^{17}$ W/Hz comes from the fact that only a few percent of small clusters, which generally correspond to less luminosity radio relics, are able to produce detectable relics, which could be seen in Figure~\ref{fig:Relic_frac}. 
}

\edited{In Figure~\ref{fig:geo_dis} we plot the distribution of the size (LLS) and the position ($R_{\mathrm{relic}}$) for the simulated radio relics with the power larger than $10^{20}$ W/Hz at 1400 MHz that are generated by all the electrons (thermal electrons and fossil electrons), and solely by thermal electrons, respectively. We exclude the relics with lower luminosity because many of them are relics stacking at $R_{\mathrm{sh,max}}$, losing most of their power after a long time of radiation cooling. 
It is obvious that there are more relics at larger radius when the fossil electrons are included, which is consistent with the fact that many relics are observed at large distance to the cluster center, and this could come from the smaller shock speed at large radius (Figure~\ref{fig:shock_evol}), which allows the shocks stay longer in this region and be more likely to be observed. On the other hand, when only thermal electrons are considered, the lower gas density and weak magnetic field at cluster outskirts make it less likely to host detectable radio relics, and the $R_{\mathrm{relic}}$ is generally evenly distributed.
}

\begin{figure*}
\includegraphics[width=\columnwidth]{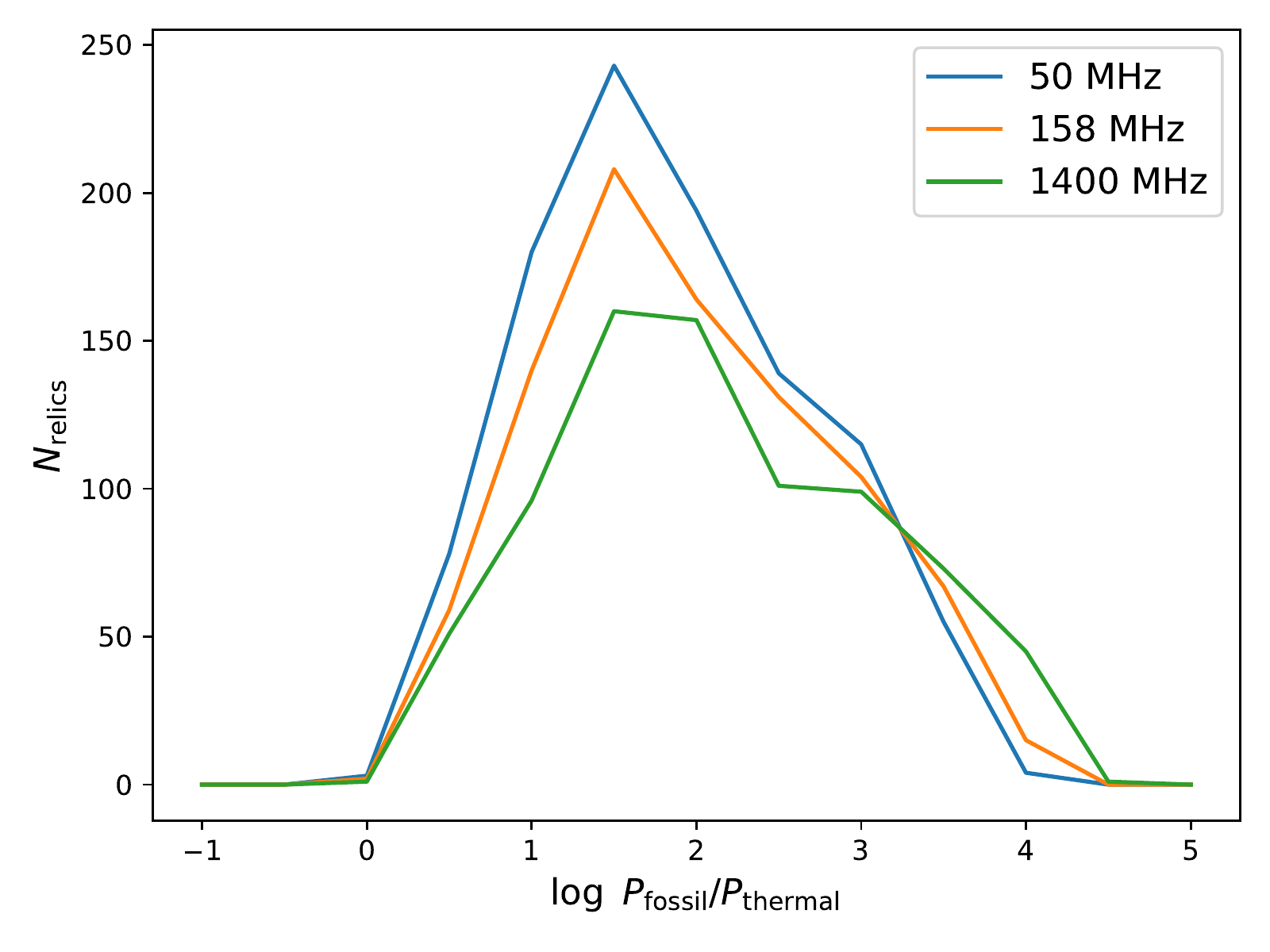}
\includegraphics[width=\columnwidth]{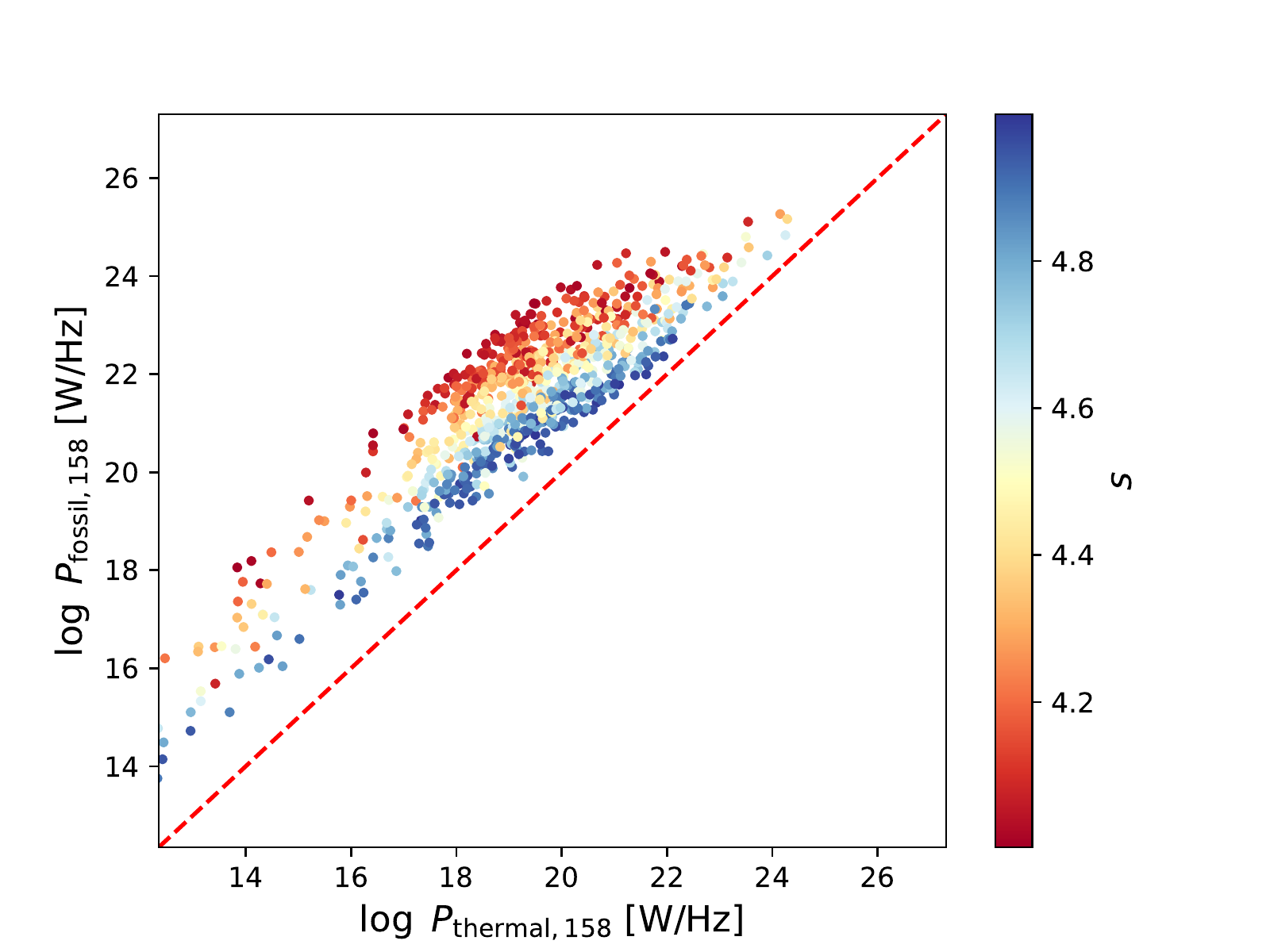}
\caption{(a) Distributions of the simulated relics as a function of the ratio between the power generated by the fossil electrons to that by the thermal electrons
(b) Distribution of the simulated relics as measured with $P_{\mathrm{fossil,158}}$ and $P_{\mathrm{thermal,158}}$. The red dash line indicates where $P_{\mathrm{fossil,158}} = P_{\mathrm{thermal,158}}$. The color represents the value of the spectral index of initial fossil electrons, i.e., $s$.
}\label{fig:fossil2thermal}
\end{figure*}

In Figure~\ref{fig:fossil2thermal} (a) we plot the relic distribution as a function of the ratio between $P_{\mathrm{fossil}}$ and $P_{\mathrm{thermal}}$ (the power produced exclusively by thermal or fossil electrons respectively) at 50 MHz, 158 MHz, and 1.4 GHz. \edited{$P_{\mathrm{fossil}}$ is larger than $P_{\mathrm{thermal}}$ by about} one order of magnitude for most relics, while in some cases the ratio can reach \edited{four} orders of magnitude.
In Figure~\ref{fig:fossil2thermal} (b) we show $P_{\mathrm{fossil,158\,MHz}}$ versus $P_{\mathrm{thermal,158\,MHz}}$, which is remarkably dependent on the spectral index $-s$, and this, again indicates that for radio relics reaccelerated fossil relativistic electrons are an important source of the radio radiation. 

\begin{figure}
\includegraphics[width=\columnwidth]{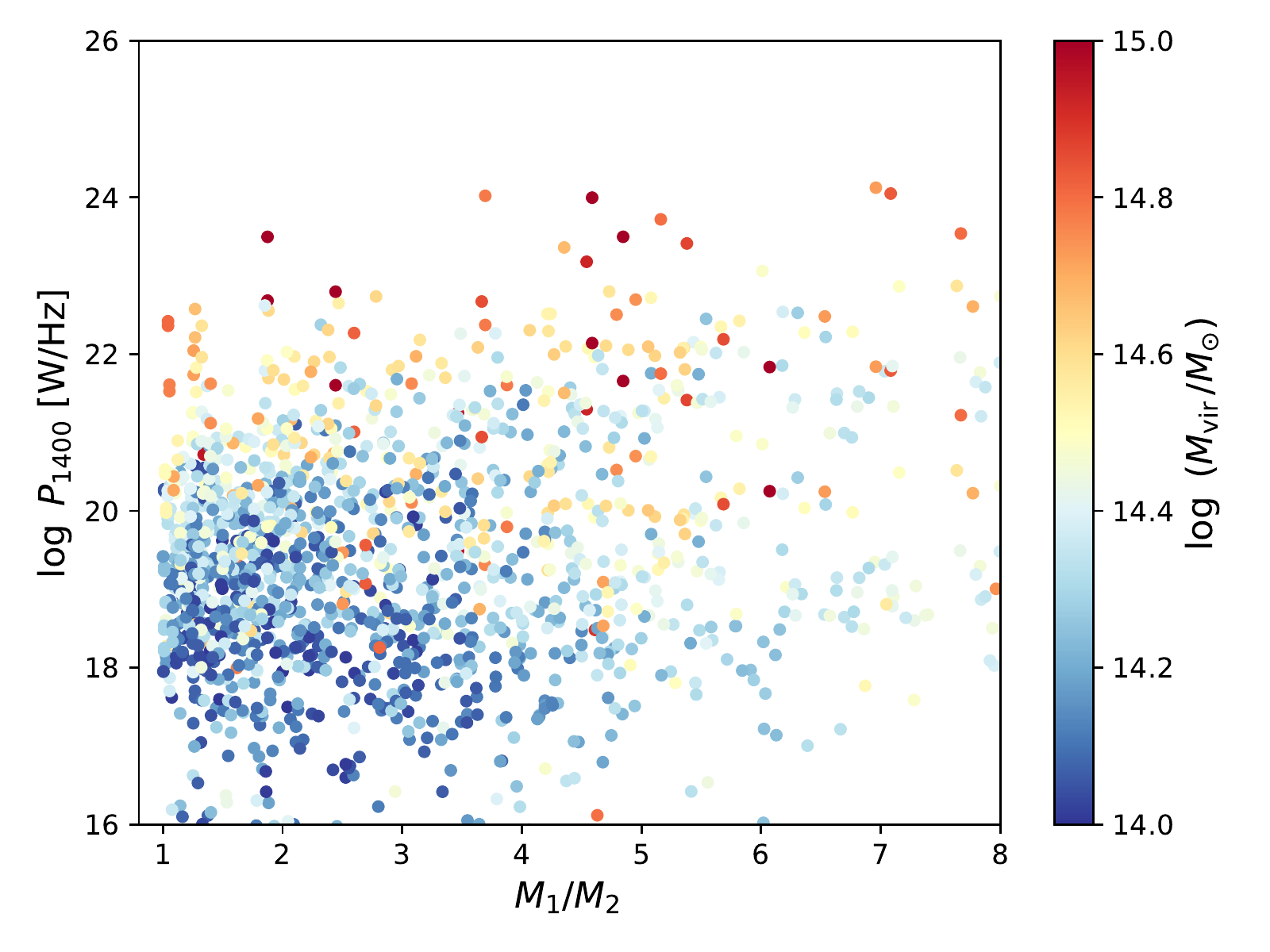}
\caption{1.4 GHz radio power versus the mass ratio $M_{1}/M_{2}$ ($M_1 > M_2$).
The virial masses of the clusters are marked with different colors.
\label{fig:P_massratio}}
\end{figure}

In Figure~\ref{fig:P_massratio} we show the power at 1.4 GHz ($P_{1400}$) versus the mass ratio of each merger, where the color represents the total mass ($M_1+M_2$). Although the scatter is large, it still can be found that 
\edited{a larger fraction of relics that hosted by minor merger system ($M_{1}/M_{2} > 3$) are powerful ones ($ P_{\mathrm{1400}} > 10^{20}$ W/Hz).}
which is consistent with the results of \citet{Vazza2020}, who suggested that unlike radio halos, radio relics are not limited to be in the major merger systems \citep{Buote_2001}. This is not unexpected since radio halos are caused by radio synchrotron emission of electrons accelerated by merger-induced turbulence, which must be more powerful in major mergers.

\section{Discussion}
\label{section:discussion}

\subsection{Origin of the Fossil Relativistic Electrons}
\label{section:CRe_sources}

\begin{figure}
\includegraphics[width=\columnwidth]{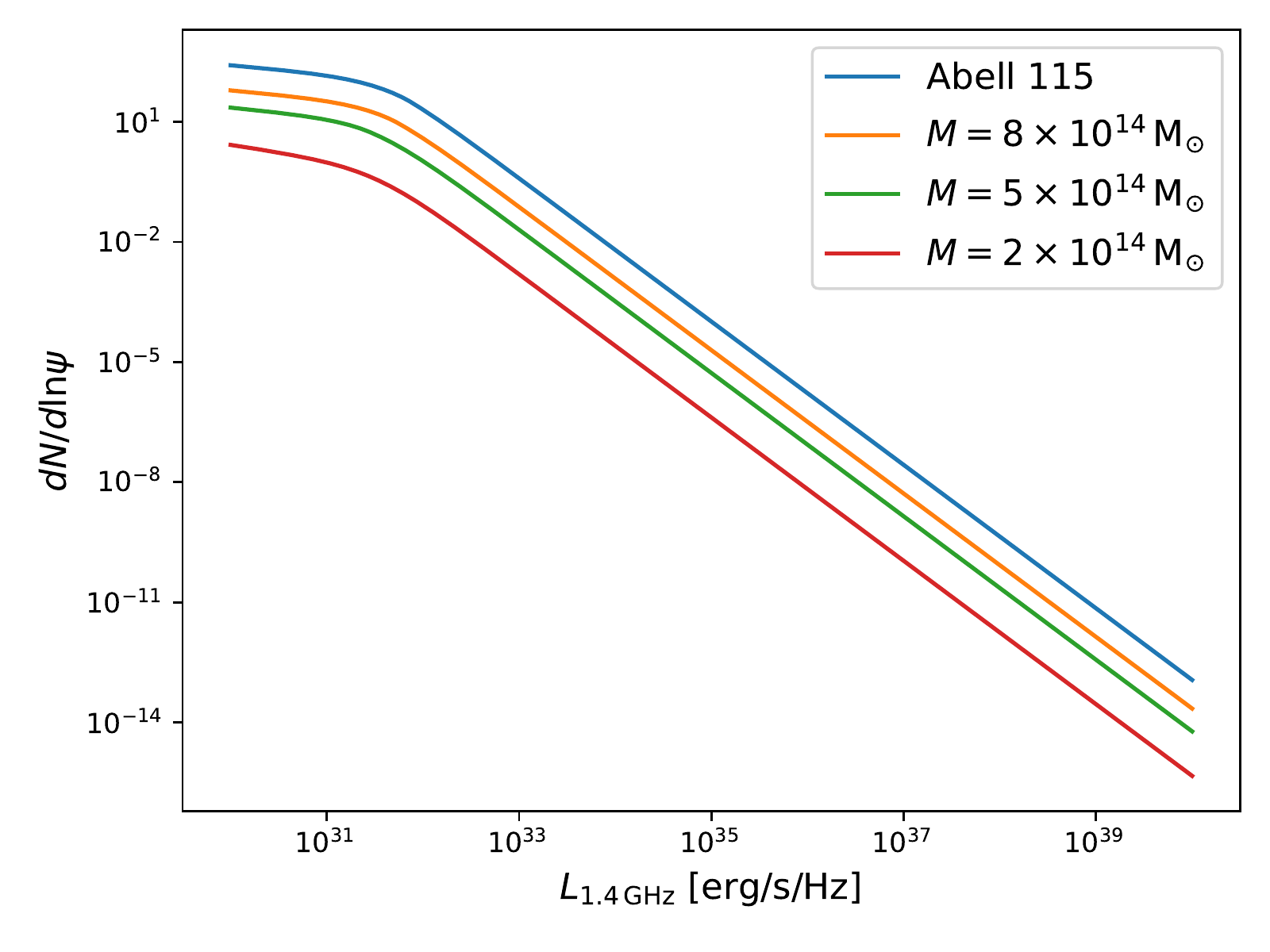}
\caption{Numbers of AGNs with different 1.4 GHz luminosities that a shock may encounter during its propagation from $0.5R_{\mathrm{vir}}$ to $1.2R_{\mathrm{vir}}$, assuming different cluster masses (colors). The case of Abell 115 is also shown (blue).
}\label{fig:AGN_prob}
\end{figure}

The fossil relativistic electrons may have several possible origins, which include previous merger-caused shock acceleration, turbulence acceleration, and AGN injection \citep{Kang2017}, as found in Abell 3411-3412 \citep{VanWeeren2017_AGN}. For example, \citet{zuhone2021b,zuhone2021A} revealed in their simulations that the CRe from AGN jets are likely transported to large radii ($\sim 0.6$ Mpc) and form bubbles with morphologies similar to radio relics.

Here we focus only on the contribution of the AGN by presenting a quantitative estimation although other possibilities are also worthy of carefully investigations. 
We calculate the probability of a merger-induced shock encountering active AGNs during their outward propagation by modeling the galaxy-AGN distribution in clusters. 
First, using the NFW model with a concentration parameter $c=3.59$ \citep{Hennig2017} to describe the density profile of cluster and assuming only one galaxy is formed in each subhalo, the number distribution of member galaxies as a function of subhalo's mass in a cluster is given by \citet{Jinag2016_halostat}
\begin{equation}
dN/d\ln{\psi}=\gamma\,\psi^{\alpha}\,\text{exp}(-\beta\,\psi^{\omega})
\end{equation}
where $\psi \equiv m/M_{H}$ is the ratio between the mass of subhalos $m$ and the mass of cluster halo $M_{\mathrm{H}}$, and $\gamma = 0.22,\ \alpha=-0.91,\ \beta = 6.00,\ \omega = 3.00$ are the parameters constrained by the observations.
Since dwarf-to-giant ratio (DGR) of galaxies does not change significantly with radius in the cluster \citep{Kopylova2013_DGR}, it is reasonable to assume that the spatial distribution of galaxies are uncoupled with galaxies' mass. 
On the other hand, by investigating the properties of 2215 galaxies each hosting a radio-loud AGN, \citet{Best2005} found that the fraction of galaxies showing AGN activity depends on the mass of the galaxy and can be described by a broken power-law
\begin{equation}\label{AGN_fraction}
f_{\text {radio-loud }}=A\left(\frac{m_{\mathrm{BH}}}{10^{8}\, \text{M}_{\odot}}\right)^{\alpha}\left[\left(\frac{L_{\mathrm{1.4\, GHz}}}{L_{*}}\right)^{\beta}+\left(\frac{L_{\mathrm{1.4\, GHz}}}{L_{*}}\right)^{\gamma}\right]^{-1}
\end{equation}
where $L_{\mathrm{1.4\, GHz}}$ is AGN's luminosity observed at 1.4 GHz, and the best fit parameter are $A = 0.0055\pm 0.0004,\ \beta = 0.35 \pm 0.03,\ \gamma = 1.54\pm 0.11,\ L_{*}=(2.5\pm0.4)\times 10^{24}\ \mathrm{W}\mathrm{Hz}^{-1}$. $m_{\mathrm{BH}}$ is the mass of the black hole in the subhalo, which scales with  $m$ \citep{Aversa2015}:
\begin{equation}
m_{\mathrm{BH}}=N \times\left[\left(\frac{m}{M_{\mathrm{b}}}\right)^{\alpha}+\left(\frac{m}{M_{\mathrm{b}}}\right)^{\omega}\right]^{-1}
\end{equation}
where $\log{N}=8.0,\ \alpha=-1.10,\ \omega=-0.80$, and $M_{\mathrm{b}} = 10^{11.90}\, \text{M}_{\odot}$. With these relations it will be straightforward to estimate how many AGNs are expected at a specific radius in a cluster with a given mass. 

Next, assuming that the shock possesses $\Omega=0.22\pi$, which is the average value of the observation sample mentioned in Section~\ref{section:solid_angle}, 
for cluster with a mass of 
$2\times 10^{14}\,\text{M}_{\odot}$, $5\times 10^{14}\,\text{M}_{\odot}$, and $8\times 10^{14}\,\text{M}_{\odot}$ at $z=0.2$, we calculate the numbers of AGNs with different radio luminosities that a shock may encounter when it travels from $0.5\,R_{\mathrm{vir}}$ to $1.2\, R_{\mathrm{vir}}$, and show the results in Figure~\ref{fig:AGN_prob}, where different colors mark the predictions for different cluster mass.
The results indicate that a typical shock does have the chance to interact with the AGN activity during its propagation, especially for those with a relatively low $L_{\mathrm{1.4GHz}}$, and the more massive the cluster is, the bigger are the chances for the shock to encounter active AGNs. This matches the observation of Abell 115 very well where 
a giant radio relic is currently coincident with a radio galaxy 
(0053+26B, $L_{\mathrm{1.5GHz}} = 1.0\times 10^{32}\ \mathrm{erg}\ \mathrm{ s}^{-1} \mathrm{ Hz}^{-1}$; \citealt{Gregorini1989}).
Detailed investigation, however, requires magnetohydrodynamics (MHD) simulations of AGN jets and shock acceleration, which is beyond the scope of this paper.

\subsection{Impact on the Detection of EoR signals}
\label{section:radioarray}

\begin{table}
    \centering
    \caption{SKA1-Low and HERA Arrays}
    \label{table:radioarray_para}
    \begin{tabular}{ccc}
        \hline
          & SKA1-Low & HERA \\
        \hline
Frequency range/MHz &50$\sim$ 350 & 50$\sim$200  \\
Field of view/deg$^{2}$  & $20.77$ & $9^{\circ}\times 9^{\circ}$  \\
Survey area       & -   & $8^{\circ}\times 180^{\circ}$       \\
Sensitivity        & $5\,\mu$Jy/1000hr  & $50\,\mu$Jy/100hr   \\
Spatial resolution     & $7''$     & $11'$            \\
        \hline
\end{tabular}

\begin{justify}
Data sources: \citet{Deboer2017} for HERA and \citet{SKA_design, Zheng2020_SKAdeepfield} for SKA1-Low.
\end{justify}
\end{table}

The spatially extended low frequency radio radiation of radio relics may cause severe foreground contamination in the observations of the cosmic 21 cm signals from the EoR with next generation instruments such as the SKA \citep{SKA_design} and the Hydrogen Epoch of Reionization Array (HERA) \citep{Deboer2017}. Although galaxy clusters are treated as foreground contaminating sources, the studies have been focused on the intergalactic medium located at cluster outskirts \citep{KESHET20041119} and radio halos \citep{Li2019}), while the impact of radio relics are barely constrained \citep{Jelic2008_foreground,Gleser2008_foreground}. In this subsection we incorporate the designs of SKA1-low and HERA arrays with our radio relic model to present a estimate of the low frequency radio contamination caused by radio relics in the 21 cm observation.

\begin{figure}
\includegraphics[width=\columnwidth]{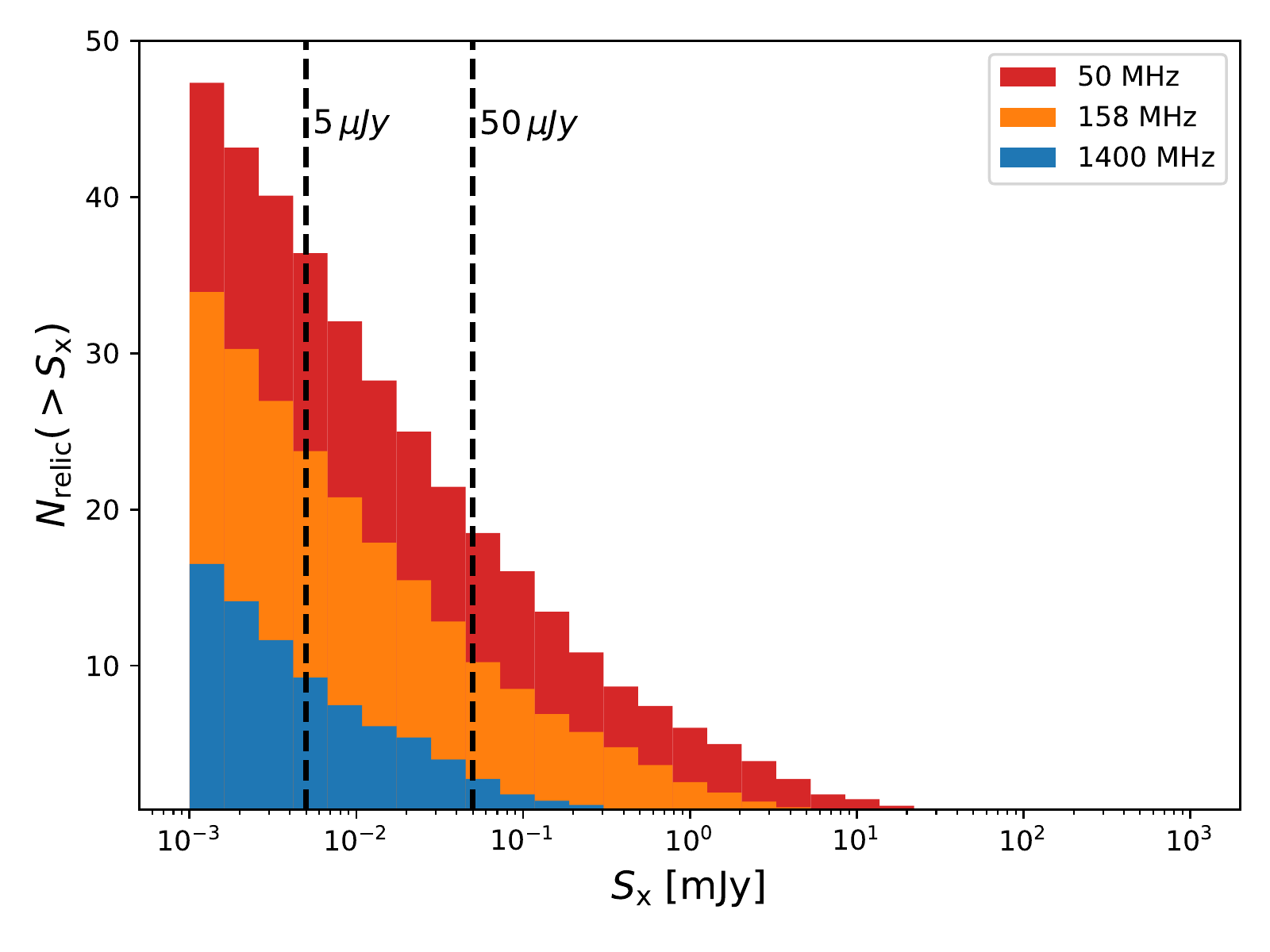}
\caption{Cumulative flux distributions of the simulated relics at 50 and 158 MHz, which are obtained \edited{by rescaling our simulation to} a $20.77\ \text{deg}^{2}$ sky patch, (i.e., the size of the planned SKA deep field). Dash lines: the SKA1-Low ($5\,\mu$Jy) and HERA ($50\,\mu$Jy) sensitivity for a 100-hour observation.
\label{fig:radio_array}}
\end{figure}

\edited{By rescaling our simulation to the required size of sky patch, whose cumulative flux distributions are shown in Figure~\ref{fig:radio_array}}, 
we find that typically \edited{34.89} (50 MHz) and \edited{22.64} (158 MHz) radio relics with $S > 5\mu $Jy will appear in the $20.77\ \text{deg}^{2}$  field of view of SKA1-low deep field. For HERA, \edited{70.86} (50 MHz) and \edited{37.87} (158 MHz) radio relics with $S > 50\mu $Jy will appear in the $9^{\circ}\times 9^{\circ}$ field of view, which are all undetectable because of its low spatial resolution ($11'$).
Their root-mean-square (rms) brightness temperature at 158 MHz calculated on a $10^{\circ} \times 10^{\circ}$ sky map with a pixel size of $20''$, together with a comparison with other contaminating sources, are summarized in Table~\ref{table:rms_Tb}. Clearly, radio relics do provided non-negligible contamination in the detection of the EoR signals, the level of which is \edited{about $1/6$ of radio halos \citep{Li2019}, and similar to galactic free-free emission, which indicates they need} to be treated very carefully in disentangling the EoR signals from the overwhelming foreground.

\begin{table}
	\centering
	\caption{The rms Brightness Temperature at 158 MHz of Several EoR Foreground Components (Unit: mK)}
	\label{table:rms_Tb}
	\renewcommand{\arraystretch}{1.25}
	\begin{tabular}{cc} 
		\hline
		Component & $T_{\mathrm{158\,MHz}}$ \\
		\hline
		Radio Relics &  \ \edited{312} \\
Radio Halos  & $1.81^{+5.28}_{-1.13} \times 10^{3}$ \\
Galactic synchrotron   & $2.52 \times 10^{5}$ \\
Galactic free-free         & 200 \\
Point sources         & $5.90 \times 10^{7}$\\
EoR signal        & 11.3 \\
		\hline
	\end{tabular}
\begin{justify}	
The $T_{\mathrm{158\,MHz}}$ of radio relics is from the simulation in this work, and results for other foreground components come from \citet{Li2019}. All of them are calculated on a $10^{\circ} \times 10^{\circ}$ sky map with a pixel size of $20''$.
\end{justify}
\end{table}

\subsection{Coulomb Collision}
\label{section:CC}

\begin{figure}
\includegraphics[width=\columnwidth]{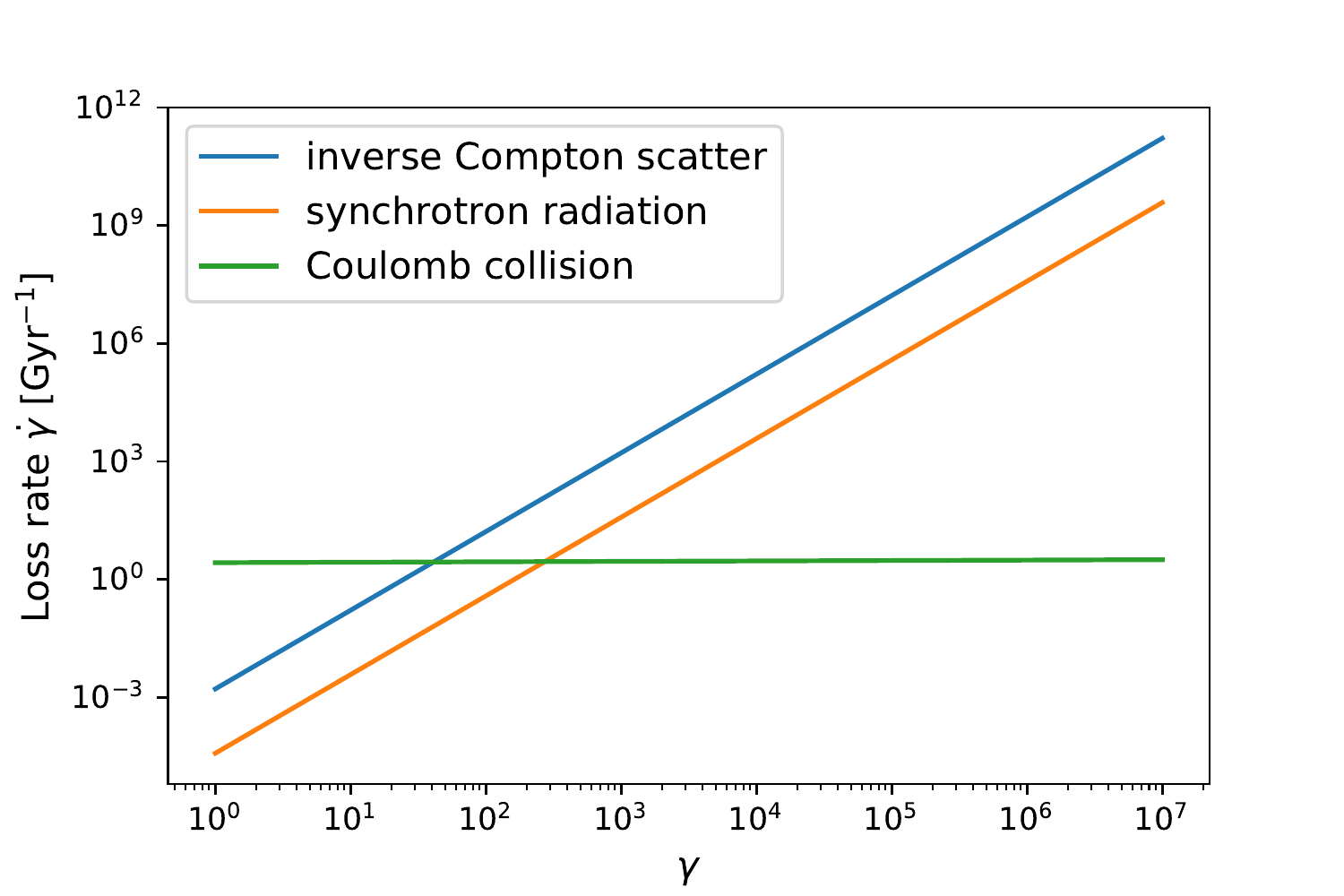}
\caption{Energy loss rates of inverse Compton scattering, synchrotron radiation, and Coulomb collision. Data are calculated for $z=0.2$ and $B=1\,\mu G$. \label{fig:loss_rate}}
\end{figure}

\begin{figure}
\includegraphics[width=\columnwidth]{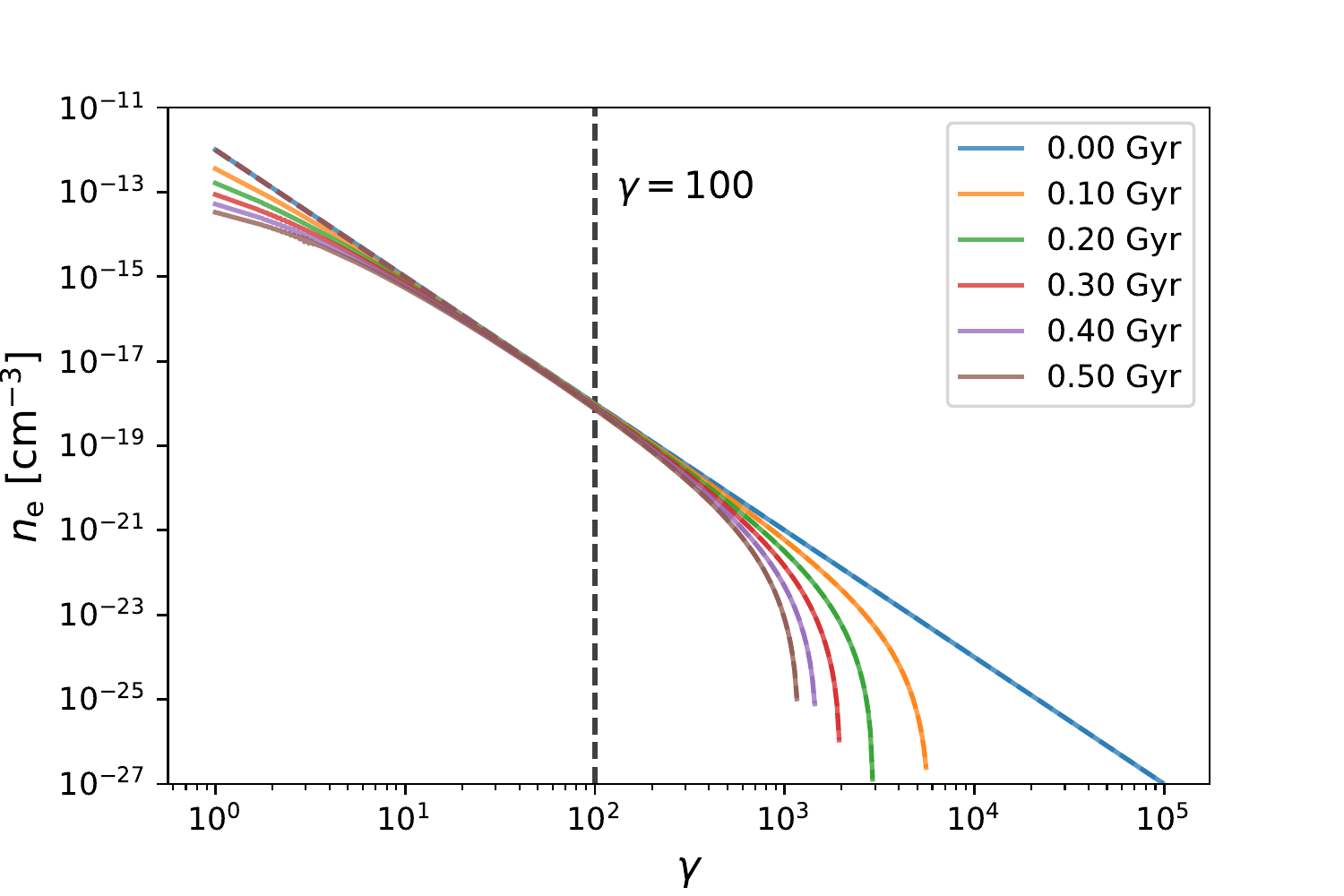}
\caption{For $z=0.2$ and $B=1\,\mu G$, evolution of the electron energy spectra when Coulomb collision is switched on (solid) or off (dash) in the calculation. 
Black dash line is the lower limit $\gamma = 100$ for $\nu \geq$ 50 MHz obtained in Section~\ref{section:CC}. 
\label{fig:CC_compare}}
\end{figure}

For a given critical frequency $\nu_{\mathrm{c}}$, the radio synchrotron emission is mainly concentrated in the frequency band of $\nu<100\nu_{\mathrm{c}}$, because the Bessel function $K_{5/3}(y)$ decreases rapidly as $y$ increases (equations~\ref{equ:syn_emit} and \ref{equ:syn_kernel}). Since $\nu_{\text{c}} = 3\,\gamma^{2}\,\nu_{\text{L}}\,\text{sin}\theta/2$, this gives the lower energy limit of the Lorentz factors $\gamma$
\begin{equation}
\gamma \geq \sqrt{\frac{\nu}{150 \nu_{\text{L}}}} \sim 4.88\times 10^{-5}\sqrt{\frac{\nu}{B}},
\end{equation}
Let $\nu_{\mathrm{min}}=50$ MHz, the typical lower limit frequency for the future radio interferometers such as SKA and HERA, we have $\gamma \gtrsim 100$ considering $B \sim \mu G$ in ICM. Incorporating these conditions with the total energy loss rate
\begin{equation}
\begin{gathered}
\left(\frac{d\gamma}{dt}\right)_{\text{total}} = -4.10 \times 10^{-5}\gamma^{2}\left(\frac{B}{1\,\mu G}\right)\\
-4.32 \times 10^{-4} \gamma^{2}(1+z)^{4}\\
-3.79 \times 10^{4}\left(\frac{n_{\mathrm{th}}}{1 \mathrm{~cm}^{-3}}\right) \times\left[1+\frac{1}{75} \ln \left(\gamma \frac{1 \mathrm{~cm}^{-3}}{n_{\mathrm{th}}}\right)\right],
\end{gathered}
\end{equation}
where the first two terms are same with those in equations~\ref{equ:syn_loss_rate} and \ref{equ:IC_loss_rate}, representing the energy loss caused by synchrotron emission and IC, and the third term corresponds to Coulomb collision (equation~\ref{equ:CC_loss_rate}), we may evaluate the impact of energy loss via Coulomb collision by switching on and off the third term. The results are summarized in Figure~\ref{fig:loss_rate}, where we show the loss rate calculated for the three energy loss mechanisms, and in Figure~\ref{fig:CC_compare}, where we show the evolution of the electrons number densities obtained by taking into account or ignoring the Coulomb collision, assuming $z=0.2$ and $B=1\,\mu G$. As can be seen in both figures, the effect of the Coulomb collision is actually not important.

\section{Conclusion}
\label{section:conclusion}
In this work we establish a semi-analytic model for radio relics by including the contribution of fossil relativistic electrons, whose properties are constrained by a sample of well-observed radio relics. 
Using Press-Schechter formalism to simulate galaxies clusters and their merger history, we obtain a radio relics catalog based on our model in a \edited{$20^{\circ} \times 20^{\circ}$} sky patch at 50 MHz, 158 MHz, and 1.4 GHz, with which 
the observed $P_{1400}-M_{\mathrm{vir}}$ relation can be successfully reconciled. 
\edited{We predict that $9.6\%$ and $7.1\%$ clusters with $M_{\mathrm{vir}} > 1.2\times 10^{14}\,\mathrm{M}_{\odot}$ would host one or more relics at 50 MHz and 158 MHz, respectively, which are consistent with the result of $10 \pm 6\%$ given by the LoTSS DR2, whose target band is 120-168 MHz.}
We explore the probability of AGNs providing seed relativistic electrons for shocks to form radio relics by calculating how many radio-loud AGNs are expected to be encountered by the shock during its propagation from $0.5\ R_{\mathrm{vir}}$ and $1.2\ R_{\mathrm{vir}}$. 
By comparing the rms brightness temperature of radio relics with those of other EoR foreground components, we find that radio relics are severe contaminating sources to EoR observations which needs serious treatment in future experiments. 
We also show the effect of the Coulomb collision is not important in the calculation of the emission of radio relics. 

\section*{Acknowledgements}
We would like to thank Congyao Zhang for helpful discussions \edited{and thank an anonymous referee for very constructive suggestions that helped to significantly improve the presentation of the paper}. This work was supported by the Ministry of Science and Technology of China (grant Nos. 2020SKA0110201, 2018YFA0404601, 2020SKA0110102), the National Natural Science Foundation of China (grant Nos. 12233005, 11973033, 11835009, 12073078, 11621303, U1531248, U1831205).

\section*{Data Availability}

The data underlying this article will be shared on reasonable request to the corresponding author.
 



\bibliographystyle{mnras}
\bibliography{ref} 








\bsp	
\label{lastpage}
\end{document}